\documentclass[12pt,oneside,english]{article}

\usepackage{amsmath}
 \usepackage{graphicx}
\usepackage{babel}
 \usepackage{setspace}
\usepackage{rotating}

\makeatletter

\bigskip \bigskip 

\begin{document}

 \centerline{\bf \Large  Gravitational brainwaves,  quantum fluctuations} 
 \centerline{\bf \Large  and stochastic quantization}

\bigskip \bigskip \bigskip  \bigskip

 \centerline{\large \bf  D. Bar}
 
 \bigskip \bigskip

\begin{abstract}   \noindent 
{\it    It is known that the biological activity of the brain involves  
radiation   of 
electric waves.  These waves result from ionic currents and charges traveling
among the brain's neurons.   
 But it is obvious that these ions and   
charges are carried by their relevant  
  masses which should give rise, according to the gravitational theory,
to extremely weak gravitational waves. 
 We use in the following the stochastic quantization
(SQ) theory to calculate the probability to find  a large ensemble of brains
radiating similar gravitational waves. We also use this SQ theory to derive
the equilibrium state related to   the known Lamb shift.  }     
\end{abstract}

\bigskip \bigskip

\noindent
{\bf \underline{Pacs numbers}}: 04.30.-w,  05.10.Gg,  02.50.Fz,  42.50.Lc

\markright{INTRODUCTION}

\bigskip \noindent 
\protect \section{Introduction \label{sec1}}
\noindent

As known, the human brain radiates, during its biological activity, several
kinds of electric waves (EW) which are generally classified as the
$\alpha$, $\beta$, $\delta$ and $\theta$ waves \cite{freeman,tran} 
(see also the references in \cite{freeman}). These EW, which differ in their
frequencies ($Hz$) and amplitudes ($\mu V$) and  are detected by electrodes 
attached to the 
scalp,  are tracked to the human states \cite{freeman} such as relaxation (related to the
$\alpha$ waves), alertness (related to the $\beta$ waves) and sleep 
which gives
rise to the $\delta$ and $\theta$ waves. The source of these  EW are 
the neurons in the cerebral cortex which are transactional cells which receive
and transmit among them inputs and outputs in the form of ionic electric
currents over short and long distances within the brain (see Chapter 1 in
\cite{freeman}). These ionic electric currents are, of course, electric charges 
in motion which may be calculated through  
 the known Gauss law \cite{halliday}.  That is,  assuming  the brain is surrounded by some hypothetical
surface $S$  one may measure, using the mentioned electrodes,  
 the electric field which crosses that surface so that he can  calculate,  
using Gauss law \cite{halliday} $ \oint {\bf E_c} \cdot d{\bf
s}=C_cq=\frac{q}{\epsilon_0} $, the charge $q$
inside the brain which is related to the measured EW.  
The  ${\bf E_c}$ in the former Gauss's law is the electric field vector and 
$\epsilon_0$ is 
the permittivity constant. 
But as known, any ion and any charge $q$ has a mass $m$  which actually carries
 it so that one may use
the corresponding Gauss's law for gravitation (see P. 618 in \cite{halliday})
$\oint {\bf E_g} \cdot d{\bf s}=C_gm=- 4\pi G\cdot m$  to relate the mass $m$ 
to
the gravitational field vector ${\bf E_g}$ which is identified at the 
neighbourhood of 
the earth 
surface  with the
gravitational acceleration  ${\bf g}$, i. e.,  ${\bf E_g}={\bf g}$.  
The constant
$G$   is the universal gravitational constant and the
 gravitational field vector at the earth surface $E_g$ is a specific 
 case of the
 generalized   gravitational waves (GW) which have tensorial properties
 \cite{mtw,hartle,thorne}. 
 These  GW are very much weak compared to the 
 corresponding EW  as may be
seen by comparing (in the  MKS system) 
 the constants which multiply the mass $m$ and charge $q$ in 
  the former two Gauss's laws, e.g, $\biggl|\frac{C_g}{C_c}\biggr|=
  \frac{4\pi G}{\frac{1}{\epsilon_0}}=4\pi \cdot 
  6.672\cdot
  10^{-11}\frac{Nm^2}{kg^2}\cdot 8.854\cdot 10^{-12}\frac{C^2}{Nm^2}=
  7.4234\cdot 10^{-25}\frac{C^2}{kg^2}$.  \par
   One may, however, consider the
real situation in which the mentioned  GW's originate not from one human 
brain but from
a large ensemble of them. Thus, if these waves have the same wavelength and
phase they may constructively interfere \cite{bar1} with each other to produce a resultant 
significant GW.    It has been shown \cite{bar1}, comparing gravitational 
waves with the 
electromagnetic ones,  that the former may also display  constructive or
destructive interference as well as holographic properties. \par 
We emphasize here before anything else that this work is not about 
consciousness,
mind or thinking  at
all (the way discussed, for example,  by Roger Penrose in his books 
\cite{penrose} or in \cite{arx})  but use only
the assumption that the mass, associated with the charge in the brain, should be
involved with gravitational field as all masses do. But, in contrast to the
electromagnetic waves,   
no GW of any kind and form were directly detected up to now, 
except through indirect
methods \cite{note},   even with the
large terrestrial interferometric Ligo \cite{ligo}, Virgo
\cite{virgo}, Geo \cite{geo} and Tama \cite{tama} detectors. Morover, 
   in contrast to other physical waves (for
example, the electromagnetic waves), GW's  do not propagate as three-dimensional (3D) 
oscillations in the background of the stationary four-dimensional (4D) 
spacetime but are themselves perturbations of this spacetime itself
\cite{mtw,hartle,thorne}. That is,   
the geometry of spacetime curves and oscillates in consequence of the presence
of the passing GW so that, in case it is strong enough, it may even impose its 
 own geometry upon the traversed spacetime \cite{beig}. 
Thus,  the  GW is an inherent part of the involved 4D spacetime  in
the sense that its geometry is reflected  
  in the related metric form 
 $ds^2$. This is seen, for example, in the metric form of the 
  cylindrical
 spacetime \cite{einstein,kuchar}  or  in the
 linearized version of general relativity where one uses the flat Minkowsky 
 metric form  to which   a small perturbation is added which
  denotes the appropriate weak passing GW \cite{mtw}.  \par 
   No one asks in such
 cases if these 4D perturbations,  which propagate as GW's, 
  occur in the background
 of some stationary higher dimensional  neighbourhood. One may, however, argue 
 that
 as other physical waves, such as the electromagnetic ones, are considered as 3D
 oscillations in the background  of the stationary 4D spacetime so the GW's may
 also be discussed as 4D oscillations in the background of a stationary 5D
 neighbourhood. This point of view was taken in the known Kaluza's 5D theory 
 and
 in the projective field formulations of general relativity 
 (unified field theories, see Chapter XVII in \cite{bergmann})
 where it was shown that the related expressions in the 5D spacetime   were 
  decomposed not only to the known Einstein field equations but also to the 
  not
  less known Maxwell equations. \par
 In this work we discuss GW from this point of view and use the stochastic
 quantization (SQ) of Paris-Wu-Namiki \cite{parisi,namiki} which is known to
 yield by a unique limiting process the equilibrium  state of many
 classical and quantum phenomena \cite{namiki}. An important and central element
 of the SQ is the assumption of an extra dimension termed in \cite{namiki}
 fictitious time in which some stochastic process, governed by either the
 Langevin \cite{coffey} or the Fokker-Plank \cite{risken} equations, is 
 performed. Thus, one may begin  from either one of the
 two mentioned  equations,  which govern the assumed stochastic process 
 in the extra dimension, and ends,  by a limiting  process in 
 which all the different 
 values of the relevant extra variable (denoted $s$)  are equated to each 
 other and taken to infinity
 \cite{namiki}, 
 in the equilibrium state.  The main  purpose of the SQ theory \cite{namiki} 
 is to obtain the expectation
 value of some random quantity or the correlation function of its variables.  
  \par  
In this work we consider, as an example of stochastic process which may be
discussed in the framework of the Parisi-Wu-Namiki SQ, the mentioned 
  activity of the human brain. 
That is,  as it is possible to calculate the correlation between a 
large ensemble of
brains in the sense of finding them radiating similar EW's  
so  one may, theoretically, 
discuss the probability 
 to  find them  radiating similar GW's. We show that although,
as mentioned, 
the GW radiated by one brain is negligible compared to the related EW the
correlation between the GW's radiated from a large number of them
 may not be small. But
in order to be able to properly calculate  this correlation we should discuss
some specific kind, from a possible large number of kinds, 
 of GW's. Thus,   we particularize to the cylindrical one   
and  calculate the probability (correlation) to find an 
ensemble of $n$ human brains  radiating 
cylindrical GW's. We do this by calculating this 
correlation in the extra
dimension and show   that once it is equated to unity one finds  that  
in the stationary state  (where the extra variable is
eliminated)  {\it  all the ensemble of brains radiate similar  
cylindrical GW's}.     
As mentioned,   no one has directly
detected, up to now,  any kind of GW 
 so all our discussion  is strictly
theoretical in the hope that some day in the future these  GW may at
last be directly detected. \par
As mentioned, the  SQ theory is suitable for discussing stochastic and
unpredictable phenomena which should be analyzed  by correlation terminology 
and probability
terms. Thus, we found it convenient to discuss the 
electron-photon interaction which originates from quantum fluctuations and 
results  in the known Lamb shift \cite{lamb} by the SQ methods. We first
calculate the states of the electron and photon and the interaction between them
in the extra dimension and then show that in the limit of eliminating the extra
variable one obtains the known expressions which characterize the Lamb shift
\cite{lamb,haken}.  
 \par  In Appendix A we represent the formalism and main expressions of the
 Parisi-Wu-Namiki SQ theory. We, especially, introduce the expressions for the
 correlation among an ensemble of variables along given intervals of the time
 $t$ 
 and the 
 extra variable  $s$. In our
 discussion here of the cylindrical GW we use the fact
 emphasized in \cite{kuchar} that {\it the ADM canonical formalism for the
 cylindrical GW  is completely equivalent to the 
  parametrized canonical formalism for the cylindrically symmetric massless
  scalar field on a Minkowskian spacetime background}. Moreover, as also
  emphasized in \cite{kuchar}, one may use the half-parametrized formalism of
  the mentioned canonical formalism without losing any important content. 
  Thus,   in Section II
 we introduce  a short review  
  of this half parametrized  cylindrical 
 massless scalar field  in the
 background of the Minkowsky spacetime where use is made of the results in
 \cite{kuchar}.  
 In Section III we represent  and discuss   
   the cylindrical GW in the framework of
  the SQ formalism and introduce  the probability that a 
  large ensemble of brains
 are found to  radiate cylindrical GW's. This probability is
 calculated in a detailed manner in Appendix $B$. In Section IV we realize  
 that the somewhat  complex expression of the 
  calculated  probability  in the extra dimension 
  is greatly simplified at the mentioned stationary limit so that one may clearly
  see that for a unity value of it all the $n$-brain ensemble radiate 
  the same 
  cylindrical GW's. 
  In Section V we discuss 
  the  electron-photon  interaction,  which results in the
  known 
 Lamb shift  \cite{lamb,haken},  in the framework of 
 the SQ formalism and the Fokker-Plank equation \cite{risken}. 
 In Section VI we   show 
   that  at the limit of the  stationary  state,   in which the extra variable is
   eliminated,  one may obtain the  
       known  expressions related to the mentioned Lamb shift as 
       obtained in the framework  of 
     quantum field
theory  \cite{lamb,haken}.    In 
Section  VII we summarize  the discussion.    
    \bigskip 
  \pagestyle{myheadings}

\bigskip

 \bigskip 
  \pagestyle{myheadings}
\markright{ THE MASSLESS CYLINDRICAL  WAVE IN THE MINKOWSKIAN ......}
       
\noindent

\protect \section{The massless cylindrical  
wave in the Minkowskian background 
 \label{sec3}}

As discussed in Appendix A the stochastic process in the extra dimension $s$ is
described by the $n$ variables $\psi(s,t)=\biggl(\psi_0(s,t),\psi_1(s,t),\ldots 
\psi_{(n-2)}(s,t),
 \psi_{(n-1)}(s,t)\biggr)$ where the finite intervals $(s_{(0)},s)$, 
  $(t_{(0)},t)$   of $s$ and $t$ during
 which the former process "evolutes" are assumed each to be subdivided 
  into $N$ subintervals   
  $(t_{(0)},t_1), (t_1,t_2), 
\ldots (t_{(N-1)},t)$ and   
$(s_{(0)},s_1), (s_1,s_2), \ldots (s_{(N-1)},s)$. In the application  of the
SQ formalism  for the ensemble of brains we identify the mentioned 
ensemble of $n$ variables $\psi_i(s,t),\ \ 0 \leq i \leq (n-1)$, 
which describe the stochastic process in the extra
dimension $s$, with the ensemble of brains. This ensemble of variables (brains)
is related, as is customary in the SQ theory, to the corresponding ensemble of
random forces $\eta(s,t)=\biggl(\eta_0(s,t),\eta_1(s,t),\ldots 
\eta_{(n-2)}(s,t),
 \eta_{(n-1)}(s,t)\biggr)$. \par
As mentioned, our aim is to calculate   
 the correlation between  the $n$-member ensemble of brains with respect to
the cylindrical GW. That is, according to the results of  
 Appendix B, 
we calculate 
the  conditional probability to find this ensemble of brains radiating 
at $t$ and $s$  the cylindrical GW's $\psi(s,t)$  
  if they were found at $t_{(N-1)}$ and $s_{(N-1)}$ 
 radiating  the cylindrical  GW's $\psi(s_{(N)},t_{(N)})$ and 
 at $t_{(N-3)}$ and $s_{(N-3)}$ they were found 
 radiating  the cylindrical  GW's $\psi(s_{(N-2)},t_{(N-2)})$ $\ldots \ldots$
 and at  $t_{(0)}$ and $s_{(0)}$ they were 
 radiating  the cylindrical  GW's $\psi(s_{(1)},t_{(1)})$ (see the discussion
 after Eqs (\ref{$B_{10}$}),   (\ref{$B_{13}$}) and  (\ref{$B_{14}$}) in
 Appendix $B$).     
   As mentioned, the cylindrical GW, 
in
its ADM canonical formalism \cite{adm}, is completetly equivalent \cite{kuchar} 
to the parametrized
canonical formalism for the cylindrically symmetric massless scalar field in 
a Minkowskian
background.  Thus,  for introducing the relevant  expressions
related to the cylindrical GW \cite{kuchar} we  write  the action
functional $S$ 
for the massless cylindrical wave in the Minkowskian
 background  \cite{kuchar,adm} 
 \begin{equation}  S=2\pi\int_{-\infty}^{\infty}dT\int_{(0)}^{\infty}dR{\cal L},
\label{e1}  \end{equation} 
 where ${\cal L}$ is is the Lagragian density  \cite{kuchar}
 \begin{equation} {\cal L} = \frac{1}{2}R\biggl((\psi_{,T})^2-(\psi_{,R})^2\biggr)
\label{e2}  \end{equation} 
The $T$ denotes the Minkowskian time and $R$ is the radial distance from the
symmetry axis in flat space \cite{kuchar}. The expressions $\psi_{,T}$,
and $\psi_{,R}$ denote the respective derivatives of $\psi$ with respect to $T$
and $R$. In the parametrized canonical formalism in a Minkowskian
background one have to  introduce \cite{kuchar} 
 curvilinear coordinates
$t$ and $r$ in flat space 
\begin{eqnarray} && t=t(T,R), \ \ \ \ r=r(T,R) \label{e3} \\ 
&& T=T(t,r), \ \ \ \ R=R(t,r) \nonumber 
\end{eqnarray}
As shown in \cite{kuchar} one  may discuss the cylindrical scalar waves in a
half-parametrized canonical formalism without losing any physical content 
except for the
spatial covariance of the scalar wave formalism \cite{kuchar}. In this
half-parametrized canonical formalism one use the following coordinates 
\begin{equation} r=R, \ \ \ \ \ t=t(T,R) \label{e4} \end{equation} 
It was shown in \cite{kuchar}, using Eqs (\ref{e1})-(\ref{e2}) and (\ref{e4}), 
 that the action $S$ assumes the simplified form 
 \begin{equation}
 S=2\pi\int_{-\infty}^{\infty}dt\int_{(0)}^{\infty}dR{\cal L}=
 2\pi\int_{-\infty}^{\infty}dt\int_{(0)}^{\infty}dR\biggl(\Pi_TT_{,t}+
 \pi_{\psi}\psi_{,t}-N{\cal H}\biggl), \label{e5} \end{equation}
 where  $T_{,t}$ and $\psi_{,t}$ denote derivatives of $T$ and $\psi$ 
 with respect to $t$.  The  $N$ is a Lagrange multiplier and ${\cal H}$ is
 \cite{kuchar} 
 \begin{equation} {\cal H}=\Pi_T+
 \underline{\cal H}, \label{e6} \end{equation}
 where  $\underline{\cal H}$ and $\Pi_T$ are related as  \cite{kuchar}
 \begin{align} & \underline{\cal H}=-\Pi_T=\frac{1}{2}\bigl(1-T^2_{,R}(R,t)\bigr)^{-1}
\biggl(-iR^{-\frac{1}{2}}\frac{\delta}{\delta \psi(R,t)} - 
 R^{\frac{1}{2}}T_{,R}(R,t)\psi_{,R}(R,t)\biggr)^2 + \label{e7} \\ & + 
\frac{1}{2}R\psi^2_{,R}(R,t) = 
\frac{1}{2(1-T^2_{,R}(R,t))}\biggl(R^{-1}\pi^2_{\psi}(R,t)
-2T_{,R}(R,t)\pi_{\psi}(R,t)\psi_{,R}(R,t) + \nonumber \\ & 
+ R\psi^2_{,R}(R,t)\biggr)  
 \nonumber 
\end{align}
 The last result were obtained by using the following definition of the momentum 
operator $\pi_{\psi}(R,t)$
\begin{equation} \pi_{\psi}(R,t)=-i\frac{\delta}{\delta(\psi(R,t))} \label{e8} 
\end{equation}
From Eqs (\ref{e6})-(\ref{e7})  one realizes that ${\cal H}$ satisfies the
constraint \cite{kuchar} \begin{equation} \label{e9} {\cal H}=0 \end{equation}
Note that we do not discuss yet 
 the SQ theory with the extra dimension which will be discussed in the following
 section. Eqs (\ref{e5})-(\ref{e9})  ensure that in the framework of the half parametrized
canonical formalism  the following 
variational principle is satisfied \cite{kuchar} 
\begin{equation}  \delta S=
\delta\biggl\{2\pi\int_{-\infty}^{\infty}dt\int_{(0)}^{\infty}
dR\biggl(\Pi_TT_{,t}(R,t)+
 \pi_{\psi}(R,t)\psi_{,t}(R,t)-N{\cal H}\biggl)\biggr\}=0, \label{e10} \end{equation}
 where all variables $T$, $\Pi_T$, $\psi(R,t)$, $\pi_{\psi}(R,t)$, and $N$ may be varied
 freely \cite{kuchar}. Note that the function $\Pi_T$ 
  may be represented  as the operator \cite{kuchar} 
 $\Pi_{T}=-i\frac{\delta}{\delta(T(R,t))}$. Also, it should be remarked   
 that the  commutation relation between 
$\pi_{\psi}(R,t)$ and $\psi_{,R}(R,t)$ is zero at the same point, i.g., 
$[\psi_{,R}(R,t),\pi_{\psi}(R',t)]
=i\frac{\delta(\psi_{,R}(R,t))}{\delta \psi(R',t)}=
i\frac{d}{dR}(\frac{\delta(\psi(R,t))}{\delta \psi(R',t)})=
i\frac{d\delta(R-R')}{dR}=0$ since the $\delta$ function is 
antisymmetric so that   one have $\frac{d\delta(0)}{dR}=0$.     
The 
  wave function $\psi(R,T)$ (not in the half-parametrized formalism),  
  which is obtained as a solution of the Einstein field
  equations for the cylindrical line element, is generally represented as an
  integral over all modes $k$ \cite{macrina}   
    \begin{equation}  \label{e11}  \psi(R,T)=\int_{(0)}^{\infty} dk
J_0(kR)\bigl(A(k)e^{(ikT)}+A^*(k)e^{-(ikT)}\bigr)  
\end{equation}
where $j_0(kR)$ is the bessel function of order zero \cite{abramowitz}.  The
quantities 
$A(k), \ A^*(k)$ denote the amplitude and its complex conjugate for some
specific mode $k$.  
 Note that here one assumes, as  done in
the  literature,   $c=\hbar=1$ so that $w={\tilde k}=p$ where 
 $w$ is the frequency, ${\tilde k}$ the  wave number and $p$ the momentum of some mode. 
The  momentum $\pi_{\psi}(T,R)$, canonically conjugate to $\psi(R,T)$, 
  may be obtained 
\cite{kuchar,macrina} by solving  the
Hamilton equation \begin{equation}  \label{e12}  \frac{\partial \psi(R,T)}{\partial
t}=\{\psi(R,T),H\}, \end{equation}  where $\psi(R,T)$ is from Eq (\ref{e11}) 
 and the curly
brackets at the right denote the Poisson brackets.   The
Hamilton function $H$ is \cite{kuchar,macrina} 
\begin{equation} \label{e13}  H=\int_{(0)}^{\infty}
dr\biggl(\tilde{N}\tilde{H}+\tilde{N^1}\tilde{H}_1\biggr)  \end{equation} 
where  $\tilde{H}$ and $\tilde{H}_1$ are respectively the rescaled 
superHamiltonian and supermomentum which where given  in \cite{kuchar} 
(see Eqs
(93)-(97) and (106)-(108) in \cite{kuchar}) as 
\begin{eqnarray} && \tilde{H}=R_{,r}\Pi_T+T_{,r}\Pi_R+
\frac{1}{2}R^{-1}\pi_{\psi}^2(R,T)+\frac{1}{2}R\psi_{,r}^2(R,T) \label{e14} \\
&& \tilde{H}_1=T_{,r}\Pi_T+R_{,r}\Pi_R+\psi_{,r}(R,T)\pi_{\psi}(R,T)  \nonumber 
\end{eqnarray}
The quantities $\psi_{,r}(R,T)$, $T_{,r}$,  $R_{,r}$ 
 denote differentiation of $\psi(R,T)$, $T$, $R$ with respect to $r$ (where in the
 half-parametrized formalism $R_{,r}=1$  as realized from Eq (\ref{e4})) and 
$\Pi_T, \
\Pi_R$ are the respective momenta canonically conjugate to $T$ and $R$. The
 $\tilde{N}$  and $\tilde{N}^1$ from Eq (\ref{e13})  
 respectively denote the rescaled 
lapse and shift
function $N$, $N^1$ (see Eqs (96) in \cite{kuchar}). Thus,  
the  $\pi_{\psi}(T,R)$ in the half-parametrized formalism 
were shown \cite{kuchar} to be
\begin{align}  & 
\pi_{\psi}(T,R)=R\biggl(\frac{(1-T^2_{,R})}{T_{,t}}\psi_{,t}(R,T)+
T_{,R}\psi_{,R}(R,T)\biggr)=
iR(1-T^2_{,R})\int_{(0)}^{\infty} dk k
J_0(kR)\cdot \nonumber \\ & \cdot \biggl(A(k)e^{(ikT)} -A^*(k)e^{-(ikT)}\biggr)
-RT_{,R}\int_{(0)}^{\infty} dk kJ_1(kR)\bigl(A(k)e^{(ikT)}+ \label{e15} \\ & + 
A^*(k)e^{-(ikT)}\bigr)+ iR\int_{(0)}^{\infty} dk k
J_0(kR)(T_{,R})^2\biggl(A(k)e^{(ikT)}-A^*(k)e^{-(ikT)}\biggr)
\nonumber 
\end{align}
where $j_1(kR)$ is the first order Bessel function \cite{abramowitz} 
obtained by differentiating $j_0(kR)$ with respect to $R$, e.g.,  
$j_0(kR)_{,R}=-kj_1(kR)$.   As shown in \cite{macrina}  
  one may express, using the expression 
$\int_{(0)}^{\infty}dr'r'\int_{(0)}^{\infty}dkkJ_n(kr)J_n(kr')f(r')=f(r)$,  
    the observables
 $A(k)$ and $A^*(k)$ in terms of $\psi(R,T)$ and $\pi_{\psi}(R,T)$ as   
  \begin{align} & A(k)=
 \frac{1}{2}\int_{(0)}^{\infty}dR
 e^{-ikT}\biggl\{Rk\biggl[\psi(R,T)\biggl(J_0(kR) -
 iT_{,R}J_1(kR)\biggr)\biggr] - iJ_0(kR)\pi_{\psi}(R,T)\biggr\}
  \label{e16} \\ &   
  A^*(k)=
 \frac{1}{2}\int_{(0)}^{\infty}dR
 e^{ikT}\biggl\{Rk\biggl[\psi(R,T)\biggl(J_0(kR)+ 
 iT_{,R}J_1(kR)\biggr)\biggr] + iJ_0(kR)\pi_{\psi}(R,T)\biggr\}
 \nonumber \end{align} 
 
 \markright{THE CYLINDRICAL GW IN THE SQ FORMALISM}
\protect \section{The cylindrical GW in the SQ formalism}

We, now, discuss the cylindrical GW from the SQ point of view 
and  begin by writing   the  Langevin 
equation (\ref{$A_{1}$}) of Appendix $A$ 
 for the subintervals $(t_{(k-1)},t_k)$ and $(s_{(k-1)},s_k)$ in the following
  form  
 \cite{namiki}
\begin{equation} \frac{\psi_i^k(s)-\psi_i^{k-1}(s)}{\bigl(s_k-s_{(k-1)}\bigr)}-
K_i(\psi^{k-1}(s))=
\eta_i^k(s), 
\label{e17}
\end{equation} 
where $\frac{d\psi_i}{ds_k}\approx \frac{\psi_i^k-\psi_i^{(k-1)}}{s_k-s_{(k-1)}}$ and 
the  $\eta_i(s)$ are conditioned  as \cite{namiki}  
\begin{equation} \label{e18} <\!\eta_i(s)\!>=0,\ \ \ 
<\!\eta_i(s)\eta_j(\grave s)\!>=\left\{ \begin{array}{ll} 0 & {\rm for~} s 
\ne \grave s \\
2\alpha\delta_{ij} & {\rm for~} s=\grave s  \end{array} \right. \end{equation} 
Note that although the $s$ dependence is emphasized in the last two equations 
one should remember that there exist also spatial and time dependence (see the
following discussion and Eq (\ref{e19})).  
 The $\alpha$  in Eq (\ref{e18}) is as discussed after Eq (\ref{$A_{3}$}) 
 of Appendix $A$. 
  The appropriate $K_i$ for the massless cylindrical scalar wave in the 
 Minkowskian
 background  may be obtained  by using Eq (\ref{$A_{2}$}) in Appendix A and Eq 
   (\ref{e5})   
  from which one realizes that the Lagrangian ${\cal L}$ depends upon two
  independent variables $t$, $R$ and five dependent varables $\psi(R,t)$,
  $\pi_{\psi}(R,t)$,  $T(R,t)$, $\Pi_T$, $N$. Note that in the following we
  represent $\psi$ and $\pi_{\psi}$ by the expressions from Eqs (\ref{e11}) and
  (\ref{e15}) as mentioned after Eq (\ref{e23}). Thus, although 
  the functions $\psi$ and $\pi_{\psi}$   should be
  denoted, because of that,  as $\psi(R,T)$ and $\pi_{\psi}(R,T)$ we denote them
  as $\psi(R,t)$ and $\pi_{\psi}(R,t)$  and take, of course, into account the
  dependence of $T$ upon $r$ and $t$ as realized, for example, 
   in Eqs (\ref{e27}). The mentioned dependence of ${\cal L}$ 
   upon the 
  dependent variables include in our case, as seen from Eqs
  (\ref{e6})-(\ref{e7}) and (\ref{e15}), dependence of  ${\cal L}$  also 
  upon some
  derivatives of them, i.e., 
    $\psi_{,t}$, $\psi_{,R}$, $T_{,t}$, $T_{,R}$. 
  Thus, the involved variation of $\delta
  S$ is given by      
  \begin{align}   & \delta
  S= 2\pi\int_{-\infty}^{\infty}dt\int_{(0)}^{\infty}dR\delta {\cal L} =
  2\pi\int_{-\infty}^{\infty}dT\int_{(0)}^{\infty}dR\biggl(\frac{\partial {\cal L}}{\partial
  \psi}\delta \psi+\frac{\partial {\cal L}}{\partial
  \psi_{,R}}\delta \psi_{,R}+\frac{\partial {\cal L}}{\partial
  \psi_{,t}}\delta \psi_{,t}+ \nonumber \\ & +\frac{\partial {\cal L}}{\partial
  T}\delta T+\frac{\partial {\cal L}}{\partial
  T_{,R}}\delta T_{,R}+ \frac{\partial {\cal L}}{\partial
  T_{,t}}\delta T_{,t}+
  \frac{\partial {\cal L}}{\partial
  \pi_{\psi}}\delta \pi_{\psi}  +\frac{\partial {\cal L}}{\partial
  \Pi_T}\delta \Pi_T+\frac{\partial {\cal L}}{\partial
  N}\delta N\biggr) \label{e19} \end{align}
 As seen from Eq (\ref{e17}) we are interested in calculating the function $K_i$
 which is given by   Eqs (\ref{e23}) and  (\ref{$A_{2}$}) in
 Appendix $A$ as 
 $K_i(\psi^{k-1}(s))=  
-(\frac{\delta S_i[\psi]}{\delta \psi})_{\psi=\psi(s,t,x)}$ where the function 
$\psi$ as function of $s$ is introduced only after varying the action $S_i$ 
 as 
functional of $\psi$.    Also, in order to deal
with compact and 
simplified expressions, as done, for example, in Eqs (\ref{e19})-(\ref{e24}), 
  we do not always write  
the various functions such as $\psi$,
$\pi_{\psi}$, $T$ etc in their full dependence upon $R$ and $T$.  \par  
 We, now, should realize    that the integrand  in the last equation (\ref{e19})
  is the total
  differential  $\delta {\cal L}$, whereas we are interested in 
  $K_i(\psi^{k-1}(t_k,s_k))$ which is seen from Eqs (\ref{e23}) and 
   (\ref{$A_{2}$}) in Appendix A 
  to be equal to
  the negative variation of the action $S_i$ with respect to $\psi$.  Thus, 
  according to the definition of $S$ from Eq (\ref{e1}) $K_i(\psi^{k-1}(t_k,s_k))$ 
   should involve  the $R$ and $t$ integration
   of the negative variation of  the Lagrangian ${\cal L}$ with respect to 
  $\psi$.  That is, we should consider only the first three terms of Eq
  (\ref{e19}) which are related to $\psi$ and its derivatives. 
Thus, for calculating the variations of these derivatives we note 
 that $\delta \psi_{,t}$, $\delta \psi_{,R}$  are  the
respective differences between the
original and varied $\psi_{,t}$, $\psi_{,R}$ and,  therefore,  they 
 may be written  as (see P. 493 in \cite{schiff})
 \begin{equation} \label{e20}  \delta \psi_{,t}=
 \frac{\partial (\delta \psi)}{\partial t}, \ \ \
 \delta \psi_{,R}=\frac{\partial (\delta \psi)}{\partial R} \end{equation}
 Using the former discussion and the last equations (\ref{e20}) 
 one may write the appropriate expression for $\delta S$ as
 \begin{equation} \label{e21} \delta
  S= \frac{\delta
  S}{\delta \psi}\delta \psi=2\pi\int_{-\infty}^{\infty}dt\int_{(0)}^{\infty}dR
  \biggl(\frac{\partial {\cal L}}{\partial
  \psi}\delta \psi+\frac{\partial {\cal L}}{\partial
 (\psi_{,R})}\frac{\partial (\delta \psi)}{\partial R}+\frac{\partial {\cal L}}{\partial
  (\psi_{,t})} \frac{\partial (\delta \psi)}{\partial t} \biggr) \end{equation}
  The second term at the right hand side of  the last equation 
  may be integrated by parts with respect
  to $R$ where the resulting surface terms are assumed to vanish  because
  $\psi$ tends to zero at infinite distances \cite{kuchar}. The third term 
  at the right hand side of  Eq (\ref{e21})
  may also be integrated by parts with respect to $t$ where the boundary terms
 vanish  because of the following assumed conditions of the   variational
 principle \cite{weinstock} $\delta \psi(R,-\infty)=\delta \psi(R,+\infty)=0$.
 Thus, Eq (\ref{e21}) becomes \begin{equation} \label{e22} \delta
  S= 2\pi\int_{-\infty}^{\infty}\int_{(0)}^{\infty}\biggl(\frac{\partial {\cal L}}{\partial
  \psi}-\frac{\partial}{\partial R} \bigl(\frac{\partial {\cal L}}{\partial
  (\psi_{,R})}\bigr)-\frac{\partial}{\partial t} \bigl(\frac{\partial {\cal L}}{\partial
  (\psi_{,t})}\bigr)  \biggr)\delta \psi dt dR \end{equation}
  We note that  analogous discussion regarding the quantization of wave
  fields may be found at pages 492-493 in \cite{schiff}. 
 Thus, using the former discussion and Eq (\ref{e22}) 
  one may write the
 following expression for $K_i(\psi^{k-1}(t_k,s_k))$
 \begin{align} & K_i(\psi^{k-1}(t_k,s_k))=  
-(\frac{\delta S_i[\psi]}{\delta \psi})_{\psi=\psi(s,t,x)}= 
-2\pi\int_{-\infty}^{\infty}dt\int_{(0)}^{\infty}dR\frac{\delta{\cal L}}{\delta
\psi}= \nonumber \\ & = -2\pi\int_{-\infty}^{\infty}
\int_{(0)}^{\infty}\biggl(\frac{\partial {\cal L}}{\partial
  \psi}-\frac{\partial}{\partial R} 
  \bigl(\frac{\partial {\cal L}}{\partial
  (\psi_{,R})}\bigr)-\frac{\partial}{\partial t} 
  \bigl(\frac{\partial {\cal L}}{\partial
  (\psi_{,t})}\bigr)  \biggr) dt dR \label{e23} 
  \end{align} 
 In order to obtain final calculable results we use, as mentioned,  for $\psi$ and $\pi_{\psi}$ 
 the respective expressions of Eqs (\ref{e11}) and (\ref{e15}).  Also, noting
 that $\pi_{\psi}$ from Eq (\ref{e15}) depends upon the derivatives $\psi_{,R}$,
 $\psi_{,t}$  one may use Eqs (\ref{e5})-(\ref{e7}) 
 and (\ref{e9}) 
 to calculate the three expressions in the integrand of the 
 last equation  (\ref{e23}) as follows
 \begin{align} & \frac{\partial {\cal L}}{\partial
  \psi}=0 \nonumber \\ &  
  \frac{\partial {\cal L}}{\partial (\psi_{,R})}= 
  T_{,t}\frac{\partial \Pi_T}{\partial (\psi_{,R})}+
  \psi_{,t}\frac{\partial \pi_{\psi}}{\partial (\psi_{,R})}=-\frac{T_{,t}}{2(1-T^2_{,R})}
  \biggl(2R^{-1}\pi_{\psi}\frac{\partial \pi_{\psi}}{\partial
  (\psi_{,R})}-2T_{,R}\pi_{\psi}- \nonumber \\ & - 2T_{,R}\psi_{,R}\frac{\partial \pi_{\psi}}{\partial
  (\psi_{,R})} + 2R\psi_{,R}\biggr)+
  \psi_{,t}\frac{\partial \pi_{\psi}}{\partial
  (\psi_{,R})}=-\frac{T_{,t}}{(1-T^2_{,R})}\bigl(R\psi_{,R}-R\psi_{,R}(T_{,R})^2\bigr)+\nonumber \\ & + 
  \psi_{,t}RT_{,R}
  = R\bigl(
  \psi_{,t}T_{,R}-T_{,t}\psi_{,R}\bigr) \label{e24}  \\ & 
   \frac{\partial {\cal L}}{\partial (\psi_{,t})} = 
   T_{,t}\frac{\partial \Pi_T}{\partial (\psi_{,t})}+\pi_{\psi}+
  \psi_{,t}\frac{\partial \pi_{\psi}}{\partial (\psi_{,t})}=-\frac{
  T_{,t}}{2(1-T^2_{,R})}\biggl(2R^{-1}\pi_{\psi}\frac{\partial \pi_{\psi}}
  {\partial
  (\psi_{,t})}-\nonumber \\ & - 2T_{,R}\psi_{,R}\frac{\partial \pi_{\psi}}
  {\partial
  (\psi_{,t})}\biggr)+  
  \pi_{\psi}+\psi_{,t}\frac{\partial \pi_{\psi}}{\partial (\psi_{,t})}=
  -\biggl(R\bigl(\frac{(1-T^2_{,R})}{T_{,t}}\psi_{,t}+T_{,R}\psi_{,R}\bigl)-
 \nonumber \\ & - 
 RT_{,R}\psi_{,R}\biggr)+ 2\frac{R(1-T^2_{,R})}{T_{,t}}\psi_{,t}
 +RT_{,R}\psi_{,R} =\frac{R(1-T^2_{,R})}{T_{,t}}\psi_{,t}
 +RT_{,R}\psi_{,R}  
  \nonumber 
  \end{align}
As seen from Eq (\ref{e23}) the expressions  
$\frac{\partial {\cal L}}{\partial (\psi_{,R})}$ and   
$ \frac{\partial {\cal L}}{\partial (\psi_{,t})}$ should be respectively
differentiated with respect to $R$ and $t$. Thus, taking into account that these
derivatives serve as integrands of integrals over $R$ and $t$ and
using  Eqs (\ref{e24}) one may
write Eq (\ref{e23}) as 
  \begin{align} & K_i(\psi^{k-1}(t_k,s_k))=   
 2\pi\biggl\{\int_{-\infty}^{\infty}
\biggl(\int_{(0)}^{\infty}\frac{\partial}{\partial R} 
  \bigl(\frac{\partial {\cal L}}{\partial
  (\psi_{,R})}\bigr)dR\biggr)dt+\int_{0}^{\infty}
\biggl(\int_{-\infty}^{\infty}\frac{\partial}{\partial t} 
  \bigl(\frac{\partial {\cal L}}{\partial
  (\psi_{,t})}\bigr)  dt \biggr) dR \biggr\}= \nonumber \\ &
  = 2\pi\biggl\{\int_{-\infty}^{\infty} dt
  \bigl(\frac{\partial {\cal L}}{\partial
  (\psi_{,R})}\bigr)\biggl|^{R=\infty}_{R=0}+\int_{0}^{\infty}dR
  \bigl(\frac{\partial {\cal L}}{\partial
  (\psi_{,t})}\bigr)\biggl|^{t=\infty}_{t=-\infty}\biggr\}= 
  2\pi\biggl\{\int_{-\infty}^{\infty} dt \biggl( R\bigl(
  \psi_{,t}T_{,R}- \nonumber \\ & 
  - T_{,t}\psi_{,R}\bigr)\biggr)\biggl|^{R=\infty}_{R=0}+
  \int_{R=0}^{R=\infty}dR\biggl(\frac{R(1-T^2_{,R})}{T_{,t}}\psi_{,t}
 +RT_{,R}\psi_{,R} \biggr)\biggr|^{t=\infty}_{t=-\infty}\biggr\} \label{e25}
  \end{align} 
  In the following   we use the  boundary values related to the
  function $T$ (see Section III in \cite{kuchar})
  \begin{equation} \lim_{t \to \pm \infty}T = t, \ \ \ 
  \lim_{r \to  \infty}T = t  \label{e26} \end{equation}
  Also, because of representing $\psi$ through 
    the expression  (\ref{e11}),  
  one may  use  the relation \cite{abramowitz}
  $j_0(kR)_{,R}=-kj_1(kR)$ in order to  write  the  derivatives of $\psi$ 
  with respect to $t$ and $r$ as
  \begin{align} &  \frac{\partial \psi(R,t)}{\partial R}=-\int_{(0)}^{\infty} dk k
J_1(kR)\bigl(A(k)e^{(ikT)}+A^*(k)e^{-(ikT)}\bigr)+ \nonumber \\ & 
+i\int_{(0)}^{\infty} dk k T_{,R}
J_0(kR)\bigl(A(k)e^{(ikT)}+A^*(k)e^{-(ikT)}\bigr) \label{e27} \\ & 
\frac{\partial \psi(R,t)}{\partial t} =
iT_{,t}\int_{(0)}^{\infty} dk k
J_0(kR)\bigl(A(k)e^{(ikT)}-A^*(k)e^{-(ikT)}\bigr) \nonumber 
 \end{align}
 Note that the leading terms of  the Bessel's functions of integer orders in 
 the
 limits of very small and very large arguments are \cite{abramowitz,schiff}
\begin{equation}  \lim_{R \to 0}J_n(R)=\frac{R^n}{(2n+1)!!}, \ \ \ \ 
  \lim_{R \to \infty}J_n(R)=\frac{1}{R}\cos\bigl(R-\frac{(n+1)\pi}{2}\bigr),  
 \label{e28}  \end{equation} 
 where $(2n+1)!!=1\cdot 3\cdot 5 \cdots (2n+1)$. 
 From the last limiting relations  one obtains for  $J_0(kR)$ and $J_1(kR)$ 
  \begin{align} &
\lim_{kR \to 0}J_0(kR)=1, \ \ \ \ \ \ \lim_{kR \to \infty}J_0(kR)= 
\frac{1}{kR}\cos\bigl(kR-\frac{\pi}{2}\bigr)\label{e29} \\ & 
\lim_{kR \to 0}J_1(kR)=\frac{kR}{1\cdot3} ,  \ \ \ \ \ \ 
\lim_{kR \to \infty}J_1(kR)=
\frac{1}{kR}\cos\bigl(kR-\pi\bigr)\nonumber \end{align}
Taking into account  Eqs (\ref{e27})   and the derivative 
$j_0(kR)_{,R}=-kj_1(kR)$
one may realize that  
    the right hand 
  side of
  Eq (\ref{e25})   becomes 
  \begin{align} & K_i(\psi^{k-1}(t_k,s_k))= 
   2\pi\int_{-\infty}^{\infty} dt\biggl(RT_{,R}\psi_{,t}- 
   RT_{,t}\psi_{,R}\biggr)
  \biggl|^{R=\infty}_{R=0}    
 + 2\pi\int_{R=0}^{R=\infty}dR\biggl(\frac{R(1-T^2_{,R})}{T_{,t}}\psi_{,t}+
 \nonumber \\ & 
 +RT_{,R}\psi_{,R} \biggr)\biggr|^{t=\infty}_{t=-\infty} 
  =2\pi\int_{-\infty}^{\infty} dt\biggl[\int_{(0)}^{\infty}dkkRT_{,t}
 J_1(kR)\biggl(A(k)e^{(ikT)}+ \nonumber \\ & + A^*(k)e^{-(ikT)}\biggr)
 \biggr]\biggr|^{R=\infty}_{R=0}
  +   2\pi\int_{0}^{\infty} dR\biggl[i\int_{(0)}^{\infty}dkkR
 J_0(kR)\biggl(A(k)e^{(ikT)}- \label{e30} \\ & - A^*(k)e^{-(ikT)}\biggr)- RT_{,R}
 \int_{0}^{\infty} dkk
 J_1(kR)\biggl(A(k)e^{(ikT)}+A^*(k)e^{-(ikT)}\biggr)\biggr]
 \biggr|^{t=\infty}_{t=-\infty}
 \nonumber  \end{align}
 Using, now, (1)  the  limiting 
relations from Eqs (\ref{e26}) and (\ref{e28})-(\ref{e29}), (2) the basic
complex relation $i^2=-1$, (3) the 
trigonometric
  identity $2i\sin(\phi)=(e^{i\phi}-e^{-i\phi})$
and (4) the   general property of Bessel's functions
of integer orders \cite{abramowitz} $\frac{d(x^nJ_n(x))}{dx}=x^nJ_{n-1}(x)$,
which reduces, for $n=1$,    to $\frac{d(xJ_1(x))}{dx}=xJ_{0}(x)$
 it is possible to show that the first two terms at the right hand side of 
 Eq (\ref{e30}) cancel each
 other
  \begin{align} &  2\pi\int_{-\infty}^{\infty} dt\biggl[\int_{(0)}^{\infty}dkkRT_{,t}
 J_1(kR)\biggl(A(k)e^{(ikT)}+ A^*(k)e^{-(ikT)}\biggr)
 \biggr]\biggr|^{R=\infty}_{R=0}+ \nonumber \\ & +
     2\pi\int_{0}^{\infty} dR\biggl[i\int_{(0)}^{\infty}dkkR
 J_0(kR)\biggl(A(k)e^{(ikT)}- 
 A^*(k)e^{-(ikT)}\biggr)\biggr]\biggl|^{t=\infty}_{t=-\infty}= \label{e31} \\ & = 
  4\pi\int_{0}^{\infty}dk
 \frac{\cos\bigl(kR-\pi\bigr)}{k}\sin(kt)\bigl(A(k)
 +A^*(k)\bigr)- 4\pi\int_{0}^{\infty}dk\sin(kt) \cdot 
  \nonumber \\ &
 \cdot \frac{\bigl(A(k)+A^*(k)\bigr)}{k}\int_{(0)}^{\infty}d(kR)
 \frac{d\bigl((kR)J_1(kR)\bigr)}{d(kR)}
 =  4\pi\int_{0}^{\infty}dk \frac{\sin(kt)}{k}
 cos\bigl(kR-\pi\bigr) \cdot \nonumber \\ & \cdot 
 (A(k)+A^*(k)) - 
 4\pi\int_{0}^{\infty}dk\frac{\sin(kt)}{k}
 \bigl(A(k) + A^*(k)\bigr)
 \cos\bigl(kR-\pi\bigr) 
=0,  \nonumber \end{align}
where we have   passed in the last result from the integral variable $R$  
to
$kR$  and use the relation from Eqs (\ref{e29})  $J_1(kR)\biggl|_{kR=0}=
\lim_{kR \to 0}J_1(kR)=
\lim_{kR \to 0}\frac{kR}{1\cdot3} \approx 0$. 
Thus, one remains with only the last term at the right hand side of Eq
(\ref{e30}) which, using Eqs (\ref{e11}), (\ref{e26}), (\ref{e29}) and 
  the integrals \cite{abramowitz}
 $\int_{(0)}^{\infty}xJ_1(x)dx=-xJ_0\biggl|^{\infty}_0+\int_{(0)}^{\infty}J_0(x)dx$ and 
 $ \int_{(0)}^{\infty}J_0(x)dx=1$,  may be reduced to  
 \begin{align} 
 & K_i(\psi^{k-1}(t_k,s_k))=- 2\pi\int_{(0)}^{\infty}dk\int_{(0)}^{\infty}d(kR)
 kR\frac{J_1(kR)}{k}T_{,R}
  \biggl(A(k)e^{ikT}+A^*(k)e^{-ikT}\biggr)\biggl|^{t=\infty}_{t=-\infty}=
  \nonumber \\ & =
 -2\pi
 \int_{(0)}^{\infty}dk
 \biggl(A(k)e^{ikt}+A^*(k)e^{-ikt}- 
 A(k)e^{-ikt}-A^*(k)e^{ikt}\biggr)T_{,R}  
 \biggl( \int_{(0)}^{\infty}d(kR)J_0(kR) - \nonumber \\ & 
 -kRJ_0(kR)\biggl|^{kR=\infty}_{kR=0}\biggr) 
= -4i\pi\int_{(0)}^{\infty}dk\sin(kt)\bigl(A(k)-A^*(k)\bigr)T_{,R}
 +\label{e32} \\ &
 +2\pi
 \int_{(0)}^{\infty}dk
 \biggl(A(k)e^{ikt}+A^*(k)e^{-ikt}-
 A(k)e^{-ikt}-A^*(k)e^{ikt}\biggr)T_{,R}kRJ_0(kR)\biggl|^{kR=\infty}_{kR=0} =\nonumber \\ &= 
 -4i\pi\int_{(0)}^{\infty}dk\sin(kt)\bigl(A(k)-A^*(k)\bigr)T_{,R}+
 2\pi\lim_{kR \to \infty}
 kR\psi(t,R)T_{,R}- \nonumber \\ & - \int_{(0)}^{\infty}dk\cos\bigl(kR-\frac{\pi}{2}\bigr) 
 \biggl( A(k)e^{-ikt}+A^*(k)e^{ikt}\biggr)T_{,R}
 \nonumber \end{align}
  We note,  as emphasized in \cite{kuchar}, that a hypersurface of constant time
  $t$ is not assumed to have conical singularity on the axis of symmetry $R=0$.
  This requires the condition \cite{kuchar} $T_{,R}=0, \  for \ R=0$. But spacetime is assumed
  to be locally Euclidean at spatial infinity \cite{kuchar} which means that the hypersurface
  of constant time $t$ have no conical singularity also at infinity so 
  that $\lim_{R \to \infty}T_{,R} \approx 0$. Thus,  one may suppose 
   that the relation  
 $\lim_{kR \to \infty}kRT_{,R}$  at Eq (\ref{e32}) tends to finite value so that
 the prefix of $\lim_{kR \to \infty}$ may be omitted.  
 It may be realized in this respect  
  from the definition of $T$ and its $r$ derivative, i.e.,
  $T(r)=T(\infty)+\int_{\infty}^r(-\pi_{\gamma})dr$, $T_{,r}=\pi_{\gamma}$ (see
  Eqs (98) and (100)  in \cite{kuchar}) that the $r$ dependence of $T$ is 
  especially
  through the $r$ at the upper end   of the integral interval.  Thus, 
   the $T_{,R}$ may be  taken outside the integral over $kR$.   
  Also, one may note 
 that the boundary
 value of $kRJ_0(kR)$ at $kR=0$ is ignored since, as seen from Eqs (\ref{e29}), 
  it obviously vanishes.  
Substituting from Eq (\ref{e32}) 
  into the Langevin equation (\ref{e17}) one
obtains 
\begin{align}   &  
\frac{\psi_{(i)}^{(k)}(s_k,R,t_k)-\psi_{(i)}^{(k-1)}\bigl(s_{(k-1)},R,t_{(k-1)}\bigr)}
{\bigl(s_k-s_{(k-1)}\bigr)} 
+4i\pi\int_{(0)}^{\infty}dk\sin(kt)\bigl(A(k)-A^*(k)\bigr)T_{,R}- \label{e33} \\ & -
2\pi \biggl[(\lim_{kR \to \infty}
 kRT_{,R})\psi(t,R)-\int_{(0)}^{\infty}dk\cos\bigl(kR-\frac{\pi}{2}\bigr) 
 \biggl( A(k)e^{-ikt}-A^*(k)e^{ikt}\biggr) T_{,R} \biggr]=
\eta_i^k   \nonumber
\end{align} 
 Thus, the  probability from Eq 
(\ref{$A_{10}$}) of Appendix A  for the subintervals 
$(t_{(k-1)},t_k)$, $(s_{(k-1)},s_k)$  
assumes the following form for the cylindrical gravitational wave 
\cite{namiki}  \begin{align}  & 
P\bigl(\psi_{(n-1)}^{(k)},t_k,s_k|\psi_{(0)}^{(k-1)},t_{(k-1)},s_{(k-1)}\bigr)=
\biggl(\frac{1}{\sqrt{2\pi(2\alpha)}}\biggr)^{n} 
\exp\biggl\{-\sum_i\biggl[\frac{1}{2(2\alpha)}
\biggl\{\frac{\bigl(\psi_{(i)}^{(k)}-\psi_{(i)}^{(k-1)}\bigr)}{\bigl(s_k-s_{(k-1)}\bigr)}+
\nonumber \\ & 
+ 4i\pi\int_{(0)}^{\infty}\sin(kt)\bigl(A(k)-A^*(k)\bigr)T_{,R} -
2\pi \biggl[(\lim_{kR \to \infty}
 kRT_{,R})\psi(t,R)- \label{e34} \\ & - 
  \int_{(0)}^{\infty}dk\cos\bigl(kR-\frac{\pi}{2}\bigr) 
 \biggl( A(k)e^{-ikt}-A^*(k)e^{ikt}\biggr)T_{R}\biggr]  \biggr\}^2\biggr]
\biggr\}, \nonumber 
\end{align}
which is the probability that the $\eta_i^k$ from the right hand side of 
Eq (\ref{e33}) takes the 
value at its left hand side \cite{namiki} and the index $i$ runs over 
the $n$ members of the
ensemble.   
 Here, we
relate the variable $s$ to the possible geometries of the gravitational wave  
in the sense that different values of $s$ refer to different geometries of the
radiated GW's.   
This is the meaning of saying that the right hand side of Eq (\ref{e33}), which
represents the  unpredictability of the stochastic forces, should 
reflects the left hand side of it which represents the variable character 
of the waves   
  radiated by the brain. 
A Markov process \cite{kannan}    in which $\eta(s)$ does not
correlate with its history is always assumed for these correlations. Eq
(\ref{e34}) is, actually, a conditional  probability which is detaily discussed
in the following section and, especially,  in Appendix $B$.
   
   \markright{THE PROBABILITY THAT THE LARGE ENSEMBLE OF BRAINS......}
   
\protect \section{The probability that the large ensemble of brains radiate
cylindrical gravitational waves}   
    The correlation for the $n$-ensemble of variables $\psi_i, \ (n-1) \geq i
    \geq 0$  over 
     the entire $N$ subintervals into which  each of the
    $(s_{(0)}, s)$
    and $(t_{(0)}, t)$ intervals are subdivided  
 may be taken from either Eq (\ref{$A_{11}$}) or the equivalent 
Eq (\ref{$A_{12}$}) of Appendix $A$  which  is  
\cite{namiki} \begin{align}  
& P\bigl(\psi_{(n-1)},t,s|\psi_{0},t_{(0)},s_{0}\bigr)=\int \cdots \int \cdots 
\int \cdots \nonumber \\ & 
\cdots P\bigl(\psi^{(N)}_{(n-1)},t_N,s_N|\psi^{(N-1)}_{(0)},
t_{(N-1)},s_{(N-1)}\bigr) \cdots  
P\bigl(\psi^{(k)}_{(n-1)},t_k,s_k|\psi^{(k-1)}_{0},t_{(k-1)},s_{(k-1)}\bigr) 
 \cdots \nonumber \\ & \cdots 
 P\bigl(\psi^{(1)}_{(n-1)},t_1,s_1|\psi^{(0)}_{0},t_{(0)},s_{0}\bigr)d\psi^{(N)}\cdots
 d\psi^{(k)} 
 \cdots d\psi^{(0)},  
 \label{e35}   \end{align} 
 where each $P$ at the right hand side of the last equation is essentially 
 given by Eq (\ref{e34}).  
In order to be able to solve the integrals in the last equation we should  
substitute
from Eq (\ref{e34}) for the $P$'s. But we should remark 
 that in Appendix $B$ and in this section 
 the relevant probability is calculated by performing the relevant summations
 first over  the $n$ variables denoted by the suffix $i$ and  then  over the
 $N$ subintervals denoted by the superscript $k$. That is, as emphasized after  
 Eq (\ref{$B_{1}$}) in Appendix $B$,  the sum over $i$ in the exponent of 
 that equation,  
 in contrast to  Eq
 (\ref{$A_{11}$}) in Appendix $A$,  precedes that over $k$ and,
 therefore, the squared expression involves the variables $\psi_{(i)}^{(k)}$, 
 $\psi_{(i-1)}^{(k)}$ etc  (instead of $\psi_{(i)}^{(k)}$, 
 $\psi_{(i)}^{(k-1)}$  of (\ref{$A_{11}$}) and Eq (\ref{e34})).   
 Now, before proceeding    we define  the
following  expressions  
 \begin{align} & B_1(R,t)=2\pi kRT_{,R} \nonumber \\ & 
 B_2(R,t)=2\pi\int_{(0)}^{\infty}dk\cos\bigl(kR-\frac{\pi}{2}\bigr) 
 \biggl( A(k)e^{-ikt}-A^*(k)e^{ikt}\biggr)T_{,R} \label{e36} \\ & 
B_3(R,t)=i4\pi\int_{(0)}^{\infty}dk\sin(kt)\bigl(A(k)-A^*(k)\bigr)T_{,R}, 
\nonumber \end{align}
where, as remarked after Eq (\ref{e32}), the prefix of $\lim_{kR \to \infty}$
were omitted from the definition of $B_1(R,t)$.  
Thus, Eq (\ref{e33}) may be written as   
   \begin{equation}  
\frac{\partial \psi(s_k,R,t)}{\partial s}
=   B_1(R,t)\psi(R,t) - B_2(R,t)-iB_3(R,t)+\eta_i^k, 
\label{e37}
 \end{equation}
  where the    
$\eta^{(k)}_i$ satisfies the Gaussian constraints from Eq (\ref{e18}).   
Solving Eq (\ref{e37}) for $\psi^{(k)}_i(s_k,R,t)$ one obtains 
\begin{align}  & \psi_i^k(s_k,R,t)=\psi_{(0)}
\exp\bigl(2\pi s_kB_{1}(R,t)\bigr)  +
2\pi\int_0^{s_k}ds'_{k}\biggl\{\exp\biggl(B_1(R,t)(s_k-s'_k)\biggr)
 \cdot \nonumber \\ &  \cdot 
\biggl(
\eta^k_i - B_2(R,t)- 
iB_3(R,t)\biggr)\biggr\}, \label{e38}  
\end{align}
for initial condition $\psi(0)=\psi_{(0)}$ at $s_k=0$. Note that differentiating Eq
(\ref{e38}) with respect to $s_k$, using the rules for evaluating integrals
dependent on a parameter \cite{pipes},  one obtains Eq (\ref{e37}). 
In Appendix $B$ we have derived in a detailed manner the appropriate expressions
for the correlations of the ensemble of $n$ variables  over the given
subintervals. We note,  as
emphasized  at the beginning of Section II, that these variables are 
related with the involved ensemble of  brains.   
Thus, the
correlation of these $n$ brains over the $N$  subinterval 
$(s_{(1)}-s_{(0)}) \ldots (s_{(N)}-s_{(N-1)})$
is given by Eq (\ref{$B_{20}$}) in Appendix $B$  as 
 \begin{align} &
 P_{i,j,l,.....}\bigl(\psi^{(N)}_{(n)},s_{(N)},t_{(N)}|\psi^{(1)}_{(0)},
 s_{(0)},t_{(0)}\bigr)= \biggl(\frac{N}
 {4\pi \alpha (\Delta s)^2\sum_{k=0}^{k=(n-1)}a_1^k}\biggr)^{\frac{1}{2}}
\cdot 
\label{e39}  \\ & \cdot
\exp\biggl\{-
\frac{N}{4\alpha (\Delta s)^2\sum_{k=0}^{k=(n-1)}a_1^k}\biggl(\psi_{(n)}^{(N)}- 
 (\sqrt{a_1})^{n+1} 
\psi_{(0)}^{(N)} 
+a_2\sum_{r=0}^{r=n+1}(\sqrt{a_1})^r\biggr)^2\biggr\},        
\nonumber  \end{align}   
where $a_1$ and $a_2$  are given in Eqs (\ref{$B_{5}$})  in Appendix $B$  as 
 $a_1=(1+2\pi B_1\Delta s)^2$,  $a_2=2\pi \Delta s(B_2+iB_3)$  
 and $\Delta s$ is
 a representative $s$ subinterval from the $N$ available 
 which are all assumed to have the same length. 
The correlation of   Eq (\ref{e39})     means, as remarked in Appendix $B$,  the
 conditional probability to find  at $s=s_{(N)}$ 
  and $t=t_{(N)}$ the  variables $\psi_{(n-1)}$, $\psi_{(n-2)}, \ \ldots  
  \psi_{(1)}$ at 
  the respective states of
  $\psi_{(n)}^{(N)}$,    $\psi_{(n-1)}^{(N)}, \ \ldots \psi_{(2)}^{(N)}$   if  
   at 
  $s=s_{(N-1)}$ 
  and $t=t_{(N-1)}$ they   were found   at   
  $\psi_{(n-2)}^{(N)}$,    $\psi_{(n-3)}^{(N)}, \ \ldots \psi_{(0)}^{(N)}$
  and at  $s=s_{(N-3)}$ 
  and $t=t_{(N-3)}$ they were found at    
  $\psi_{(n-2)}^{(N-2)}$,    $\psi_{(n-3)}^{(N-2)}, \ \ldots \psi_{(0)}^{(N-2)}$
 $\ldots \ldots$ and at $s=s_{(0)}$ and $t=t_{(0)}$ they were at 
  $\psi_{(n-2)}^{(1)}$,    $\psi_{(n-3)}^{(1)}, \ \ldots \psi_{(0)}^{(1)}$. 
  That is, the conditional probability here includes a condition for each of 
    the
   $N$ subintervals $(s_{(0)}, s_{(1)}),  (s_{(2)}, s_{(3)}), \
  \ldots 
  (s_{(N-1)}, s_{(N)})$ so that  the superscripts of the variables $\psi$ 
 at the beginnings of all these subintervals  are the same as at the ends 
 of
  them as remarked after Eqs (\ref{$B_{10}$}), 
  (\ref{$B_{13}$}),   (\ref{$B_{14}$}) and (\ref{$B_{15}$}) in Appendix $B$.  
   \par
  From the last equation (\ref{e39})  one may realize that for assigning to    
  $ P_{i,j,l,.....}\bigl(\psi^{(N)}_{(n)},s_{(N)},t_{(N)}|\psi^{(1)}_{(0)}, \\
 s_{(0)},t_{(0)}\bigr)$  a probability meaning 
  which have values only in the range $(0, 1)$ the following inequality should
  be satisfied 
  \begin{align} & 
 \biggl(\frac{4\pi \alpha (\Delta s)^2\sum_{k=0}^{k=(n-1)}a_1^k}{N}\biggr)^{\frac{1}{2}}
\geq   
\exp\biggl\{
\frac{N}{4\alpha (\Delta s)^2\sum_{k=0}^{k=(n-1)}a_1^k}\biggl(
 (\sqrt{a_1})^{n+1} 
\psi_{(0)}^{(N)} - \label{e40} \\ & - \psi_{(n)}^{(N)}
-a_2\sum_{r=0}^{r=n+1}(\sqrt{a_1})^r\biggr)^2\biggr\} \nonumber 
\end{align} 
Taking the $\ln$ of the two sides of the last inequality and solving for
$\psi_{(n)}^{(N)}$ one obtains 
\begin{align} & \psi_{(n)}^{(N)} \geq (\sqrt{a_1})^{n+1} 
\psi_{(0)}^{(N)}-a_2\sum_{r=0}^{r=n+1}(\sqrt{a_1})^r- \label{e41} \\ & -
\biggl[\biggl(\frac{2 \alpha (\Delta s)^2\sum_{k=0}^{k=(n-1)}a_1^k}{N}\biggr)
\ln\biggl(\frac{4\pi \alpha (\Delta s)^2\sum_{k=0}^{k=(n-1)}a_1^k}{N}\biggr)
\biggr]^{\frac{1}{2}},
\nonumber \end{align}
where for a unity probability one should consider the equality sign of the last
inequality. That is, if the variables $\psi_{(n)}^{(N)}$ and 
 $\psi_{(0)}^{(N)}$ are related to each other in the extra dimension 
 according to the equality sign of
 (\ref{e41}) then the probability to find  at the
 equilibrium state  (where the variable $s$ is eliminated) 
 the whole ensemble of variables  all 
   related to 
 the same gravitational geometry is unity.  And since, as  remarked, these 
 variables are
 identified with the discussed ensemble of brains this means that they   
  are all radiating  cylindrical GW's. 
  This may be shown when one equates all the different values
 of $s$   to each other 
and taking  the  infinity limit as should be done in  the stationary 
configuration.  In such case one have $\Delta s=0$
and therefore it may be realized  from
Eqs (\ref{$B_{5}$}) in Appendix $B$ that the following relations are valid 
 \begin{align} & a_{1_{\Delta s=0}}=\sqrt{a_{1_{\Delta s=0}}}=
 (\sqrt{a_{1_{\Delta s=0}}})^{(n+1)}=1, \ \ \ \ 
 \sum_{r=0}^{r=n+1}(\sqrt{a_{1_{\Delta s=0}}})^r=(n+2) \label{e42} \\ &  
  \sum_{k=0}^{k=(n-1)}a_{1_{\Delta s=0}}^k=n, \ \ \ \ 
a_{2_{\Delta s=0}}=0   \nonumber \end{align}  That is, using the last 
relations and noting that the 
 $\ln$ function 
  satisfies the limiting relation \cite{pipes} 
$\lim_{x \to 0} x^2\ln(x^2)=0$ one obtains from Eq (\ref{e41}) the expected
stationary state \begin{equation} \label{e43} \psi_{(n)_{st}}^{(N)} =
\psi_{(0)_{st}}^{(N)} \end{equation}
Noting the way by which the conditional probability from Eq (\ref{$B_{20}$}) 
in Appendix
$B$ was derived and the fact that $N$ and $n$ denote general numbers 
it may be realized  that the last result from Eq (\ref{e43}) 
 ensures  that at $t=t_{(N)}$ in the equilibrium situation all the variables
$\psi_{i_{st}}^{(N)}, \ \  0 \leq i \leq n $ are equal to
each other.  This means that the  probability to find the related  ensemble of
brains all radiating  at $t_{(N)}$  cylindrical GW
$\psi_{(i)_{st}}^{(N)}$  is unity.  \par
Note from the discussion in Appendix $B$ that the stationary state  from 
Eq (\ref{e43}) 
 have been  obtained  by inserting the cylindrical GW  Langevin 
expression   from Eq (\ref{e33})-(\ref{e34})  
into  the action $S_k$  for   
each  subinterval  $  (s_{(k-1)},s_k), \ \ 1 \le k \le N$  
 of each  member of the ensemble of variables as realized from Eqs 
 (\ref{$B_{1}$})-(\ref{$B_{3}$}) in Appendix
 $B$. 
   This kind of substitution is clearly 
   seen in  Eq (\ref{e34}) which  includes the Langevin relation from
 (\ref{e33}) in each variable $\psi_i, \ 0 \leq i \leq n$ and for each
 subinterval $(s_{(k)}-s_{(k-1)}), \ 1 \leq k \leq N$. 
 As one may realize from Eqs (\ref{e41})-(\ref{e43})  the substituted expressions differ by
 $s$  and only at the limit that these expressions have the same $s$ 
  that one finds the same cylindrical GW pattern  shared by all the 
 ensemble members.      
    Thus, 
when these differences  in $s$  are eliminated by  
equating,  in the stationary state,  all the $s$ values  to each other 
 one may obtain the situation in which all the members of the
 ensemble of brains radiate  cylindrical GW  and, 
  therefore,  
 the  correlation     is maximum.

\markright{ THE ELECTRON-PHOTON INTERACTION AND STOCHASTIC....... }
         
\bigskip        \protect
\section{The  electron-photon interaction  
and
 stochastic quantization}
The main lesson one learns  from the discussion in the former sections about 
the gravitational
brainwaves is that introducing the cylindrical GW expression into the 
actions $S$
of the path integrals related to the mentioned ensemble of  variables (brains)  
results with the outcome  that the probability to
find them radiating this kind of waves  is large. In this section
we demonstrate this again regarding the quantum fluctuations which cause the
shifting of the energy bands in the known Lamb shift experiment \cite{lamb}.  
Here the
ensemble of stochastic processes do not represent, as in the previous sections, 
any biological brain activity but the action of a two-state electron which emits
a photon and then reabsorbs it where the total energy during this process is not
conserved. This process, which is tracked to  quantum fluctuations \cite{lamb},  
 is regarded here in the framework
of the SQ theory as obtained in the  equilibrium limit of some stochastic
process in an extra dimension $s$.  That is, discussing this phenomenon as a 
stochastic process occuring in an extra dimension 
  we
show that taking  the steady state limit of equating all the involved  $s$
values to each other and taking to infinity one obtains the 
 results of the Lamb shift experiment \cite{lamb}. \par
As is customary in the SQ theory and exemplified in the former sections we
assume that there exist in an extra dimension a large ensemble of 
 stochastic processes  each of them may give rise   in the stationary state to 
 the
Lamb shift phenomenon.   Also, it is assumed that each of these stochastic
processes is performed during finite $s$ and $t$
intervals $(s_{(0)}, s)$, $(t_{(0)}, t)$ and that    
 each of  these  intervals 
 is subdivided 
 into an $N$   
subintervals $(s_{(0)},s_1)$, $(s_1,s_2)$, \ldots $(s_{N-1},s_N)$ and $(t_{(0)},t_1)$, 
$(t_1,t_2)$, \ldots, $(t_{N-1},t_N)$. \par 
In the following we formulate the 
appropriate
expression for the described electron-photon interaction over some 
 representative 
subintervals  
$(t_{(k-1)},t_k)$ and $(s_{(k-1)},s_k)$ and  calculate the probability to
find the ensemble of stochastic processes giving rise to the same 
   remarked electron-photon
interaction. In contrast to the discussion in the former sections where we use
the stochastic Langevin formulation of the SQ theory 
 we, now,  find it better to discuss the 
 equivalent  
  Fokker-Plank version of it 
\cite{namiki,risken}. 
 That is, we use the following Fokker Plank equation \cite{namiki,risken}
 \begin{equation} \frac{\partial 
 P(\psi^{(k)},t_{(k)},s_{(k)}|\psi^{(k-1)},t_{(k-1)},s_{(k-1)})}
 {\partial s}=
F^{(k)}P(\psi^{(k)},t_{(k)},s_{(k)}|\psi^{(k-1)},t_{(k-1)},s_{(k-1)}),   
 \label{e44}  \end{equation} 
where  $P(\psi^{(k)},t_{(k)},s_{(k)}|\psi^{(k-1)},t_{(k-1)},s_{(k-1)})$  
 denotes  the conditional probability to find the relevant ensemble of 
 stochastic processes
 giving rise  
 at $t_k$ and $s_k$ to the state  $\psi^{(k)}$ if at the former $t_{(k-1)}$ and   
 $s_{(k-1)}$   they give rise to  the state 
$\psi^{(k-1)}$. In the context of this section  the states
$\psi^{(k)}$ and 
$\psi^{(k-1)}$ are in effect  two  total situations     
  each of them includes  all the 
particular  photon-electron interaction states related to the ensemble 
of stochastic
variables at the two different $t$ and $s$ values of 
$t_{(k)}$,    
 $s_{(k)}$  and $t_{(k-1)}$  $s_{(k-1)}$. In this way the $P$'s here have
 similar meaning to the $P$'s of the former sections which are related to
 cylindrical GW's.    
The  $F^{(k)}$ in Eq (\ref{e44}) is  \cite{namiki} \begin{equation} 
\label{e45} 
F^{(k)}= \frac{1}{2\alpha}H(\psi^{(k)},\pi^{(k)}), \end{equation} where $H$,  
$\pi^{(k)}$ and $\psi^{(k)}$ are,  respectively,  
the ``stochastic'' Hamiltonian,  
momentum and  state for the subintervals $(s_{(k-1)}, s_k)$,  
 $(t_{(k-1)}, t_k)$. The $\alpha$, as mentioned after Eq  (\ref{e3}), is either 
 $\alpha=\frac{k_{\beta}T}{f}$ for classical phenomena or $\alpha =\hbar$ for
 quantum ones. 
   The momentum  $\pi^{(k)}$ is,  as in quantum mechanics, a differential operator
   defined by   \cite{namiki} $\pi^{(k)}=
   -2\alpha \frac{\partial }{\partial 
   \psi^{(k)}}$  
and  satisfied the  commutation relations  
\cite{namiki} $[\pi^{(m)},\psi^{(n)}]= 2\alpha \delta^{mn}$. The operator $F$
from Eq (\ref{e45}) is also a differential operator which may be written 
generally for the
ensemble of $n$ stochastic processes as
\cite{namiki} 
\begin{equation}
F=\sum_{i=1}^{i=n}\biggl(\alpha\frac{\partial^2}{\partial((\psi_{(i)})^2)}-
\frac{\partial K_{(i)}(\psi)}{\partial (\psi_{(i)})}\biggr) \label{e46} \end{equation} 
Noting that $K_{(i)}(\psi)$ has the same meaning as in the Langevin formalism of the
SQ theory (see
Eq (\ref{e17}) and Eq (\ref{$A_{1}$}) in Appendix $A$) one may write 
   the last  equation 
(\ref{e46}) in a manner which emphasizes the underlying stochastic process 
$\eta$
  \begin{equation} \label{e47} F=
  \frac{\partial }{\partial (\psi_{(i)})}\sum_{i=1}^{i=n}\biggl(
  \alpha\frac{\partial}{\partial(\psi_{(i)})}-
 K_{(i)}(\psi)\biggr) = \frac{\partial }{\partial (\psi_{(i)})}\biggl\{\sum_{i=1}^{i=n}\biggl(
 \alpha\frac{\partial}{\partial(\psi_{(i)})}-
 \biggl( \frac{\partial \psi_{(i)}}{\partial
 s}-\eta_{(i)}\biggr)\biggr)\biggr\}  \end{equation} 
 As emphasized in \cite{namiki} one may
develop, using the former relations,   a stochastic operator formalism 
which corresponds to the  quantum
one so that it is possible to formulate  a  ``Schroedinger'', ``Heisenberg'' 
and
``interaction'' pictures.  
Thus, assuming an ensemble of $n$ stochastic processes, using the "interaction"
picture and considering the whole intervals   $(t_{(0)},t)$, and  $(s_{(0)}.s)$ 
one may calculate  the conditional probability
 to find at $s$ and $t$ these processes giving rise to the  state   
   $\psi$ if at the initial 
 $s_{(0)}$
 and $t_{(0)}$ they give rise to the state  $\psi^{(0)}$. This conditional probability 
 is given by  
    \cite{namiki} 
 \begin{eqnarray} && P^I(\psi,t,s|\psi^{(0)},t_{(0)},s_{(0)})=
 P^I(\psi^{(0)},t_{(0)},s_{(0)})+
 \label{e48} \\ && + \int_{(0)}^{\psi}
 F^{(N)}P^I(\psi^{(N-1)},t_{(N-1)},s_{(N-1)}|\psi^{(0)},t_{(0)},s_{(0)})
 d\psi^{(N)}, \nonumber 
  \end{eqnarray} 
 where the superscript $I$ reminds us that we use the "interaction" picture and 
 $P^I(\psi^{(0)},t_{(0)},s_{(0)})$ is the probability that the ensemble of stochastic
 processes give rise at the initial  
 $t_{(0)}$ and $s_{(0)}$ to  the  initial state $\psi^{(0)}$.   The states  
  $\psi$ depends upon $s$ and $t$ and, therefore,   the integration 
   over
 $\psi$ is, actually, a double one  over $s$ and $t$. Thus, substituting  
   in a
 perturbative manner \cite{feynman}   for 
 $P^I(\psi^{(N-1)},t_{(N-1)},s_{(N-1)}|\psi^{(0)},t_{(0)},s_{(0)})$   one may 
 write 
   Eq
 (\ref{e48})  as
  \begin{align} &
 P^I(\psi,t,s|\psi^{(0)},t_{(0)},s_{(0)})= P^I(\psi^{(0)},t_{(0)},s_{(0)})+
 \sum_{k=1}^{k=N}\int_{\psi^{(0)}}^{\psi}F^{(N)}d\psi^{(N)}
 \int_{\psi^{(0)}}^{\psi^{(N)}}F^{(N-1)}d\psi^{(N-1)} \cdot \nonumber
 \\ & \cdot
 \int_{\psi^{(0)}}^{\psi^{(N-1)}}F^{(N-2)}d\psi^{(N-2)}
 \ldots \int_{\psi^{(0)}}^{\psi^{(k)}}F^{(k-1)}d\psi^{(k-1)}  
 \ldots \int_{\psi^{(0)}}^{\psi^{(3)}}F^{(2)}d\psi^{(2)}  \cdot  \label{e49} \\ &
 \cdot 
  \int_{\psi^{(0)}}^{\psi^{(2)}}F^{(1)} d\psi^{(1)}
  P^I(\psi^{(0)},t_{(0)},s_{(0)})
    =P^I(\psi^{(0)},t_{(0)},s_{(0)})+ 
 \int_{\psi^{(0)}}^{\psi}d\psi^{(1)}F^{(1)}P^I(\psi^{(0)},t_{(0)},s_{(0)})+
 \nonumber \\ &  
 + 
 \int_{\psi^{(0)}}^{\psi}d\psi^{(2)} \int_{\psi^{(0)}}^{\psi^{(2)}}d\psi^{(1)}
 F^{(1)}
 F^{(2)} P^I(\psi^{(0)},t_{(0)},s_{(0)})+ 
  \int_{\psi^{(0)}}^{\psi}d\psi^{(N)} \int_{\psi^{(0)}}^{\psi^{(N)}}d\psi^{(N-1)} 
 \ldots  \nonumber \\ & \ldots 
\int_{\psi^{(0)}}^{\psi^{(3)}}d\psi^{(2)}\int_{\psi^{(0)}}^{\psi^{(2)}}
d\psi^{(1)}
 F^{(1)}F^{(2)}
 \ldots F^{(N-1)}F^{(N)}P^I(\psi^{(0)},t_{(0)},s_{(0)}) \nonumber 
 \end{align}
 Note that in the last equation we have obtained  in
 each term the same factor of  $P^I(\psi^{(0)},t_{(0)},s_{(0)})$.  
 Now, since the Lamb shift results from quantum fluctuations  and since the
 states in 
 quantum mechanics as  well as in  SQ \cite{namiki}  have a probabilistic  
 interpretation we may 
 assume 
 that  
 the probabilities $P^I$   denote states. We 
 should, however, emphasize (again)   
   that these
   $P$'s  from
  Eq (\ref{e49}), as those of the previous sections (see, for example, Eqs
  (\ref{e34})-(\ref{e35})), refer  to the states of the whole ensemble 
  of stochastic
  variables in the sense of the conditional probability to find them at a 
  later $s$ and $t$ 
   in some situation
  $\psi$ if, for example,  at
  the initial  $s_{(0)}$ and $t_{(0)}$ they were at the  situation
  $\psi^{(0)}$.  We later at Eqs (\ref{e50})-(\ref{e59}) denote  the respective 
  particular
  states of the interacting electron and photon by  
  $\phi$ and $u$.  \par
   Thus,  following the last discussion one may use 
 the   quantum rules and
 terms as in \cite{haken}, except for the introduction of the extra 
 variable $s$,
 for representing the electron and photon before and after the interaction
 between them as well as the general state of the whole ensemble of stochastic
 variables. 
  The variable $s$ is introduced into the relevant
 quantities so that in the limit of equating all the different $s$ values to
 each other and taking to infinity, as required in the SQ
 theory \cite{parisi,namiki},   the known expressions \cite{haken}
 which   represent the electron and photon and the correlation between them 
 are obtained.    
  Thus,
 one  may assign to the initial $s_{(0)}$ and $t_{(0)}$ 
 the value of zero
 and refer to $P^I(\psi^{(0)},t_{(0)}=0,s_{(0)}=0)$ as the initial state 
 of the  
 ensemble system.   \par 
 As remarked, the electron  is assumed to have 
 two different states so  that at $t_1$ and $s_1$ it was at the higher state 2
 from which  it descends to the lower state 1 through emitting a photon. Then at
 $t_2$ and $s_2$ it reabsorbs the photon and returns to state 2 as schematically
 shown at the left hand side of Figure 1. In the following we denote 
  the higher and lower  
 energies  of the electron by $\epsilon_2$ and $\epsilon_1$ respectively and that of 
 the photon by 
 $w_{\lambda}$ where, due to the nonconserved energy character of the
 interaction, $\epsilon_2-\epsilon_1 \neq w_{\lambda}$. We wish to represent 
 the $s$
 dependence of the electron and photon in the extra dimension in a similar
 manner as their $t$ dependence.   The conventional  $t$ dependence 
  (see, for example, Chapter 7 in \cite{haken}) of
 an incoming electron with energy $\epsilon_2$ at time $t_1$ (before any 
 interaction of it) is $e^{-i\epsilon_2t_1}$ and that of an outgoing electron 
 with energy $\epsilon_1$ at time $t_2$ (after any interaction of it) is 
 $e^{i\epsilon_1t_2}$. The  $t$ dependence of the 
 emitted photon at $t_1$ is  \cite{haken} 
 $e^{iw_{\lambda}t_1}$ and that of the  reabsorbed photon  at $t_2$ by 
 $e^{-iw_{\lambda}t_2}$. 
  Thus, according to the former discussion the $(s,t)$ dependence of 
  the incoming electron $\phi(s,t)$  and the emitted 
 photon  $u(s,t)$    
  at $t_1$ and $s_1$ 
 may be represented by 
\begin{eqnarray} && \phi(s_1,t_1)_{before \ emission}=
e^{-i\epsilon_2t_1}+e^{-i\epsilon_2s_1(1-i\delta)} \label{e50} \\ &&
  u(s_1,t_1)_{after \ emission}=e^{iw_{\lambda}t_1}+e^{iw_{\lambda}s_1(1+i\delta)},  \nonumber 
  \end{eqnarray}        
   where $\delta$ is an 
infinitesimal satisfying $\delta \cdot \infty=\infty$, and $\delta \cdot c=0$, 
 ($c$ is a constant) \cite{mattuck}. This is done   
  so that for finite values of $s$ the dependence upon $s$,  for both the electron
  and photon,  is similar, as remarked,  to the dependence upon $t$ 
  and when  $s \to \infty$, which is the equilibrium situation
   in  the SQ theory,  
   the terms in $s$
 vanish  as required. That is
  \begin{eqnarray} && \phi(s_1<\infty,t_1)_{before \ emission}=
e^{-i\epsilon_2t_1}+e^{-i\epsilon_2s_1} \nonumber \\ &&
 \lim_{s \to \infty}\phi(s,t_1)_{before \  emission}=
e^{-i\epsilon_2 t_1}
    \label{e51} \\ && 
 u(s_1<\infty,t_1)_{after \ emission}=e^{iw_{\lambda}t_1}+e^{iw_{\lambda}s_1}, \ \ \ \ \ 
 \lim_{s \to \infty}u(s,t_1)_{after \ emission}=e^{iw_{\lambda}t_1} \nonumber  
  \end{eqnarray}       
The expression for the outgoing electron at $t_1$ and $s_1$ with the lower 
energy $\epsilon_1$ 
(after emitting the photon)  and its reduction for finite and infinite $s$ 
 are  \begin{eqnarray} && \phi(s_1,t_1)_{after \ emision}=e^{i\epsilon_1t_1}+
 e^{i\epsilon_1s_1(1+i\delta)},  \label{e52} \\ && \phi(s_1<\infty,t_1)_{after \
 emission}=
e^{i\epsilon_1t_1}+e^{i\epsilon_1s_1}, \ \ \  
\lim_{s \to \infty}\phi(s,t_1)_{after \ emission}=
e^{i\epsilon_1 t_1},  \nonumber 
 \end{eqnarray}  where the 
 $\delta$ has the same meaning as 
 before. 
Just before  the reabsorption stage at $t_2$
  and $s_2$ the  electron and photon are  represented 
    by \begin{eqnarray}  && \phi(s_2,t_2)_{before \ reabsorption}=
    e^{-i\epsilon_1t_2}+e^{-i\epsilon_1s_2(1-i\delta)} \label{e53} 
    \\ && u(s_2,t_2)_{before \ reabsorption}=e^{-iw_{\lambda}t_2}+e^{-iw_{\lambda}s_2(1-i\delta)} 
    \nonumber \end{eqnarray} Needless to remark that, according to our
    discussion,  the former expressions
    reduce, for finite and infinite $s$, to  
     \begin{eqnarray} && \phi(s_2<\infty,t_2)_{before \ reabsorption}=
e^{-i\epsilon_1t_2}+e^{-i\epsilon_1s_2} \nonumber \\ &&
 \lim_{s \to \infty}\phi(s,t_2)_{before \ reabsorption}=
e^{-i\epsilon_1 t_2}
    \label{e54} \\ && 
u(s_2<\infty,t_2)_{before \ reabsorption}=e^{-iw_{\lambda}t_2}+e^{-iw_{\lambda}s_2}, \ \ \ \ \ 
 \lim_{s \to \infty}u(s,t_2)_{before \ reabsorption}=e^{-iw_{\lambda}t_2} \nonumber  
  \end{eqnarray}       
   Just   after the
    reabsorption at $s_2$ and $t_2$ the expression for the electron and its
    reduction for finite and infinite $s$ are  
    \begin{eqnarray} && \phi(s_2,t_2)_{after \ reabsorption}= 
    e^{i\epsilon_2t_2}+e^{i\epsilon_2s_2(1+i\delta)}
    \label{e55} \\ && \phi(s_2<\infty,t_2)_{after \ reabsorption}=
e^{i\epsilon_2t_2}+e^{i\epsilon_2s_2} \nonumber \\ &&
 \lim_{s \to \infty}\phi(s,t_2)_{after \ reabsorption}=
e^{i\epsilon_2 t_2} \nonumber 
    \end{eqnarray}  
     Beside the former expressions for the separate electron and photon  we should
 take into acount also the interaction between them, that is, the emission and
 reabsorption of the photon by the electron. This interaction for the emission
 part in the extra dimension $s$, denoted   $g^{em}_{\lambda_s}$,  may be
 written as
 \begin{equation}   g^{em}_{\lambda_s}=\label{e56} -\sqrt{
 \frac{e^2}{2m^2\hbar w_{\lambda}\epsilon_0}}\int \phi_1(s,t) u(s,t) p \phi_2(s,t)dV,
 \end{equation}
where $\phi_2(s,t)$, $\phi_1(s,t)$,  denote the two energy states of the electron as given by Eqs
(\ref{e50})-(\ref{e55}) and $u(s,t)$ is the expression for the photon given by Eqs
(\ref{e50})-(\ref{e51}) and (\ref{e53})-(\ref{e54}). The $w_{\lambda}$ and 
$\epsilon_0$ are respectively the energy of the emitted photon and the
dielectric constant in vacuum.   The integration is over the volume which 
includes also the $s$ dimension and the $p$ is the momentum operator which is
represented by $p=\frac{\hbar}{i}\Delta$. The former expression for the
emission interaction is suggested so that in the limit of $s \to \infty$ it
reduces to the known emission interaction which does not involve the $s$
variable (see Eq (7.112) in \cite{haken}). That is, 
\begin{align} & \lim_{s \to
\infty}g^{em}_{\lambda_s}=\lim_{s \to \infty}\biggl\{-\sqrt{
 \frac{e^2}{2m^2\hbar w_{\lambda}\epsilon_0}}\int \phi_1(s,t) u(s,t) p
 \phi_2(s,t)dV\biggr\}= \nonumber \\ & = -\sqrt{
 \frac{e^2}{2m^2\hbar w_{\lambda}\epsilon_0}}\int 
 \lim_{s \to \infty}(\phi_1(s,t)) \lim_{s \to \infty}(u(s,t)) p
 \lim_{s \to \infty}(\phi_2(s,t))dV= \label{e57} \\ & = -\sqrt{
 \frac{e^2}{2m^2\hbar w_{\lambda}\epsilon_0}}\int \phi_1(t) u(t) p
 \phi_2(t)dV=g^{em}_{\lambda} \nonumber 
 \end{align}
 where the last result is obtained by noting from  Eqs (\ref{e50})-(\ref{e55})  and 
 that in the limit $s \to \infty$ the expressions for the electron and photon
 reduce to their known forms \cite{haken}.
The interaction for the reabsorption part may be obtained  
 by noting that the expressions for  
the electron and photon participating in
the reabsorption interaction are obtained by taking  the hermitian adjoints  
of the expressions for the electron and photon participating 
in the emission process. Thus, using the rule \cite{schiff,merzbacher} that the hermitian
adjoint of the product of some expressions is the product of their adjoints in
the reverse order, one may obtain the interaction 
  for the reabsorption part, denoted $g^{re}_{\lambda_s}$, from that
  of the emission part  $g^{em}_{\lambda_s}$ as follows
  \begin{align}  & g^{re}_{\lambda_s}=
  \biggl({g^{em}_{\lambda_s}\biggr)}\dagger=\biggl({-\sqrt{
 \frac{e^2}{2m^2\hbar w_{\lambda}\epsilon_0}}\int \phi_1(s,t) u(s,t) p
 \phi_2(s,t)dV}\biggr)\dagger =  \label{e58}  \\ & = -\sqrt{
 \frac{e^2}{2m^2\hbar w_{\lambda}\epsilon_0}}
 \int ({\phi_2(s,t)})\dagger  ({p})\dagger 
  ({u(s,t)})\dagger ({\phi_1(s,t)})\dagger dV \nonumber 
  \end{align}
   The reabsorption interaction reduces  at the limit of $s \to \infty$, 
  just like the emission process in Eq
  (\ref{e57}),  to the known reabsorption
  interaction \cite{haken} which does not involve the extra $s$ variable. That is,  
   \begin{align} & \lim_{s \to
\infty}g^{re}_{\lambda_s}=\lim_{s \to \infty}\biggl\{-\sqrt{
 \frac{e^2}{2m^2\hbar w_{\lambda}\epsilon_0}}
 \int ({\phi_2(s,t)})\dagger  ({p})\dagger 
  ({u(s,t)})\dagger ({\phi_1(s,t)})\dagger dV  \biggr\}= \nonumber \\ & =
   -\sqrt{
 \frac{e^2}{2m^2\hbar w_{\lambda}\epsilon_0}}\int 
 \lim_{s \to \infty} ({\phi_2(s,t)})\dagger  p
 \lim_{s \to \infty}({u(s,t)})\dagger \lim_{s \to \infty}({\phi_1(s,t)})\dagger 
 dV= 
 \label{e59} \\ & = -\sqrt{
 \frac{e^2}{2m^2\hbar w_{\lambda}\epsilon_0}}\int ({\phi_2(t)})\dagger p
 ({u(t)})\dagger 
 ({\phi_1(t)})\dagger dV=g^{re}_{\lambda} \nonumber 
 \end{align}
 Note that the whole processes of emission and reabsorption may, respectively, 
  be 
 read directly
from Eqs (\ref{e57}) and (\ref{e59}) if one realizes that the operator 
$p$ in each of
these equations denotes the interaction undergone by
the   expressions (denoting electron or (and) photon) at its right hand side 
 which result with   the expressions 
 (also denoting electron or (and) photon) at its left hand side. Thus,
in Eq (\ref{e57}), which describes the emission process, the $\phi_2(s,t)$  at
the right of $p$ denotes the initial electron   with 
the higher energy state 2 and
the $\phi_1(s,t)u(s,t)$ 	at the left of $p$ are the electron with the 
lower
energy state 1 and the emitted  photon. Likewise, in Eq (\ref{e59}), which describes 
the reabsorption  process, the $({u(s,t)})\dagger ({\phi_1(s,t)})\dagger$  at
the right of $p$ denotes the initial lower energy electron and the photon, 
before the reabsorption,   and
the $({\phi_2(s,t)})\dagger$ 	at the left of $p$ is the electron 
with the higher 
energy state 2 after the reabsorption.

\markright{THE LAMB SHIFT AS A STATIONARY STATE OF  STOCHASTIC......}

\protect \section{The  Lamb shift as a stationary state of stochastic processes
in the extra dimension}

	 Now,   we must realize that the final
  state at $t$ and $s$ after the reabsorption of the photon,  where we remain 
  with
  one electron with the higher energy state 2,   is the same as the 
  initial state at $t_{(0)}$ and $s_{(0)}$ before the  emission of the photon 
  from the  higher energy electron. Thus, 
  we may  write for the relevant $P^I$ at
  the end of the whole  process of emission and reabsorption \cite{haken} 
  \begin{equation}
  \label{e60}  P^I(\psi,t,s|\psi^{(0)},t_{(0)},s_{(0)})=
  P^I(\psi^{(0)},t_{(0)},s_{(0)})+G(s,t)
  P^I(\psi^{(0)},t_{(0)},s_{(0)}),  
  \end{equation}  where the coefficient $G(s,t)$ denotes the mentioned evolution
  during the $(t_{(0)},t)$ and $(s_{(0)},s)$ intervals from 
   the initial state $P^I(\psi^{(0)},t_{(0)},s_{(0)})$ back
   to the same state.  We first note    that as the $(s,t)$ dependence of the 
   states of the electron and
   photon were represented  as sums of two terms, one involves only 
   the $t$ term and the second only the $s$ term, so the  $(s,t)$ 
   dependence of the entire  mentioned 
     interaction  of
  (emission$+$reabsorption) $G(s,t)$ 
   may also be written as a sum of two separate terms, denoted  
  $G(t)$ and $G(s)$   each of them involves 
   only one variable. This is done, as will just be realized, so that at the
   equilibrium limit the $s$ term vanishes and remains only the $t$ term as 
   is the case regarding the mentioned $(s,t)$ representation of the
   states of the electron and photon (see Eqs (\ref{e50})-(\ref{e55})). \par
     Thus, for the $t$ dependence of the emission process 
     one should take  into account that: (1) the emission process  
     is executed during the
     interval   $0 < t_1 < t_2$ , (2) the electron before and after  emission
    at $t_{(1)}$ is, respectively, represented by $e^{-i\epsilon_2t_1}$ 
     and $e^{i\epsilon_1t_1}$,  (3) the emitted photon at $t_{(1)}$ is given 
     by $e^{iw_{\lambda}t_1}$ and (4) the emission itself is described by the
     interaction  $g^{em}_{\lambda}$.  
      And for the $t$ dependence of the reabsorption  process 
     one should take  into account that: (1) the reabsorption 
      process 
     is executed during the
     interval   $0 < t_2 < t$ , (2) the electron before and after  
      reabsorption at $t_{(2)}$ 
     is, respectively, represented by $e^{-i\epsilon_1t_2}$ 
     and $e^{i\epsilon_2t_2}$,  (3) the reabsorbed photon at $t_{(2)}$ is given 
     by $e^{-iw_{\lambda}t_2}$ and (4) the reabsorption itself is described by the
     interaction  $g^{re}_{\lambda}=
     ({g^{em}_{\lambda}})\dagger$.      Thus, one may write the $t$
     dependence of the (emission$+$reabsorption) process $G(t)$ as
   \begin{eqnarray} &&  G(t)=
 g^{em}_{\lambda}\cdot ({g^{em}_{\lambda}})\dagger \cdot
\int_{0}^{t_2}\exp\biggl(i\bigl(\epsilon_1+w_{\lambda}-
\epsilon_2\bigr)t_1\biggr)dt_1 \cdot \label{e61}
\\   && \cdot 
\int_{0}^t\exp\biggl(i\bigl(\epsilon_2-w_{\lambda}-\epsilon_1\bigr)t_2\biggr)dt_2 \nonumber 
\end{eqnarray}
 Simiarly,  for the $s$ dependence of the emission process 
     one should take  into account that: (1) the emission process  
     is executed during the
     interval   $0 < s_1 < s_2$ , (2) the electron before and after  emission
     is, respectively, represented by $e^{-i\epsilon_2s_1(1-i\delta)}$ 
     and $e^{i\epsilon_1s_1(1+i\delta)}$,  (3) the emitted photon is given 
     by $e^{iw_{\lambda}s_1(1+i\delta)}$ and (4) the emission itself is 
     described by the
     interaction  $g^{em}_{\lambda_s}$.  
      And for the $s$ dependence of the reabsorption  process 
     one should take  into account that: (1) the reabsorption 
      process  
     is executed during the
     interval   $0 < s_2 < s$, (2) the electron before and after  reabsorption 
     is, respectively, represented by $e^{-i\epsilon_1s_2(1-i\delta)}$ 
     and $e^{i\epsilon_2s_2(1+i\delta)}$,  (3) the reabsorbed photon is given 
     by $e^{-iw_{\lambda}s_2(1-i\delta)}$  and (4) the reabsorption itself is described by the
     interaction  $g^{re}_{\lambda_s}=
     ({g^{em}_{\lambda_s}})\dagger$.      Thus, one may write the $s$
     dependence of the (emission$+$reabsorption) process $G(s)$ as
\begin{eqnarray}
  && G(s)=
 g^{em}_{\lambda_s}\cdot ({g^{em}_{\lambda_s}})\dagger \cdot
\int_{0}^{s_2}\exp\biggl[i\biggl(\epsilon_1+i\delta\bigl(\epsilon_2+\epsilon_1+
w_{\lambda}\bigr)+w_{\lambda}-\epsilon_2\biggr)s_1\biggr]ds_1 \cdot \nonumber \\
&& \cdot \int_{0}^s\exp\biggl[i\biggl(\epsilon_2+i\delta\bigl(\epsilon_2+
\epsilon_1+w_{\lambda}\bigr)-
w_{\lambda}-\epsilon_1\biggr)s_2\biggr]ds_2,   \label{e62} 
 \end{eqnarray} 
 where we have set, as remarked,  $s_{(0)}=t_{(0)}=0$ for both $G(t)$ 
 and $G(s)$.  The coefficient    
    $G(t,s)$ from Eq (\ref{e60}) is given, as remarked, 
     by  the sum $G(t)+G(s)$ so that 
    in the equilibrium
    state obtained in the limit in which all the values of $s$ are
    equated to each other and taken to infinity the term $G(s)$ vanishes 
    and remains only the term
    $G(t)$ as should be \cite{haken}. The term $G(s)$ vanishes in the stationary
    state because  we have already equated
    the initial $s_{(0)}$ to zero so for equating all the  $s$'s  to 
    each other
    one have to set also the other values of $s$ equal to zero which 
    obviously causes
    $G(s)$ from Eq (\ref{e62}) to vanish.   Note that  thus far
    we have discussed a single mode $\lambda$ for the emitted and reabsorbed 
    photon which makes sense in a cavity whose closed walls are of the same
    order as the wavelength of the photon.  But for an infinite space or 
    a cavity with open sides 
    one
    should consider a continuum of modes $\sum_{\lambda}$. 
    Thus, considering this continuum of modes and  performing the integration
    over $t_1$ and $s_1$ from Eqs (\ref{e61})-(\ref{e62}) one obtains   
 \begin{align} & G(t,s)=G(t)+Gs)=
\sum_{\lambda}g^{em}_{\lambda}\cdot ({g^{em}_{\lambda}})\dagger \cdot \int_{0}^tdt_2
\frac{\biggl\{\exp\biggl[i(\epsilon_1+w_{\lambda}-
\epsilon_2)t_2\biggr]-1\biggr\}}{i(\epsilon_1+w_{\lambda}-\epsilon_2)} \cdot
\nonumber \\ & \cdot 
\exp\biggl[i(\epsilon_2-w_{\lambda}-\epsilon_1)t_2\biggr]+
 \sum_{\lambda_s}g^{em}_{\lambda_s}\cdot ({g^{em}_{\lambda_s}})\dagger 
 \cdot \label{e63}  
 \\ & \cdot 
\int_{0}^sds_2\frac{\biggl\{\exp\biggl[i\biggl(\epsilon_1-
\epsilon_2+w_{\lambda}+i\delta(\epsilon_2+\epsilon_1+w_{\lambda})\biggr)
s_2\biggr]-1\biggr\}}
{i\biggl(\epsilon_1-\epsilon_2+w_{\lambda}+
i\delta(\epsilon_2+\epsilon_1+w_{\lambda})\biggr)} \cdot \nonumber \\ & \cdot
\exp\biggl[i\biggl(\epsilon_2- \epsilon_1-w_{\lambda}+ 
i\delta\bigl(\epsilon_2+\epsilon_1+w_{\lambda}\bigr)\biggr)s_2\biggr] \nonumber \end{align}

Now, performing the integration over $s_2$ and $t_2$ we obtain from Eq
(\ref{e63})
\begin{align}  & G(s,t)=G(t)+G(s)=
\sum_{\lambda}
\frac{g^{em}_{\lambda}\cdot ({g^{em}_{\lambda}})\dagger}{i\bigl(\epsilon_1+
w_{\lambda}-\epsilon_2\bigr)}\biggl\{t-
\frac{\exp\biggl(i\bigl(\epsilon_2-
\epsilon_1-w_{\lambda}\bigr)t\biggr)-1}{i
\bigl(\epsilon_2-\epsilon_1-w_{\lambda}\bigr)}\biggr\}
+ \nonumber \\ & +
\sum_{\lambda_s}
\frac{g^{em}_{\lambda_s}\cdot ({g^{em}_{\lambda_s}})\dagger}
{i\biggl(\epsilon_1-\epsilon_2+w_{\lambda}+
i\delta\bigl(\epsilon_2+\epsilon_1+w_{\lambda}\bigr)\biggr)}
\biggl\{-\frac{\exp\biggl(-2\delta \bigl(\epsilon_2+\epsilon_1+w_{\lambda}\bigr)s\biggr)
-1}
{2\delta\bigl(\epsilon_2+\epsilon_1+w_{\lambda}\bigr)}- \label{e64} \\ & 
-\frac{\exp\biggl[i\biggl(\epsilon_2-
\epsilon_1-w_{\lambda}+i\delta\bigl(\epsilon_2+\epsilon_1+w_{\lambda}\bigr)
\biggr)s\biggr]-1}
{i\biggl(\epsilon_2-\epsilon_1-w_{\lambda}+
i\delta\bigl(\epsilon_2+\epsilon_1+w_{\lambda}\bigr)\biggr)}\biggr\} \nonumber \end{align}
 One may realize that, because  of the $\delta$ (see its definition after Eq
 (\ref{e50})), the  quotient $-\frac{\exp\biggl(-2\delta \bigl(\epsilon_2+\epsilon_1+w_{\lambda}\bigr)s\biggr)
-1}
{2\delta\bigl(\epsilon_2+\epsilon_1+w_{\lambda}\bigr)}$ in the 
 second sum, which is of the
 kind $\frac{0}{0}$, may be evaluated,   using  L'hospital theorem
 \cite{pipes},  to obtain for it the result of $s$ so that Eq (\ref{e64}) 
  becomes
 \begin{align}  & G(s,t)=G(t)+Gs)=
\sum_{\lambda}
\frac{g^{em}_{\lambda}\cdot({g^{em}_{\lambda}})\dagger}{i\bigl(\epsilon_1+
w_{\lambda}-\epsilon_2\bigr)}\biggl\{t-
\frac{\exp\biggl(i\bigl(\epsilon_2-
\epsilon_1-w_{\lambda}\bigr)t\biggr)-1}{i
\bigl(\epsilon_2-\epsilon_1-w_{\lambda}\bigr)}\biggr\}
+ \nonumber \\ & +
\sum_{\lambda_s}
\frac{g^{em}_{\lambda_s}\cdot({g^{em}_{\lambda_s}})\dagger}
{i\biggl(\epsilon_1-\epsilon_2+w_{\lambda}+
i\delta\bigl(\epsilon_2+\epsilon_1+w_{\lambda}\bigr)\biggr)}
\biggl\{s- \label{e65} \\ & 
-\frac{\exp\biggl[i\biggl(\epsilon_2-
\epsilon_1-w_{\lambda}+i\delta\bigl(\epsilon_2+\epsilon_1+w_{\lambda}\bigr)
\biggr)s\biggr]-1}
{i\biggl(\epsilon_2-\epsilon_1-w_{\lambda}+
i\delta\bigl(\epsilon_2+\epsilon_1+w_{\lambda}\bigr)\biggr)}\biggr\} \nonumber \end{align}
 The last expression for $G(t,s)$ contains  terms 
which  are proportional to $t$ and
$s$,   others which  are oscillatory in these variables,  and also 
constant terms. Thus,  
 for large $t$ and $s$ the oscillatory as well as the constant terms may
be neglected compared to $t$ and $s$ as in the analogous quantum discussion  
of the
same process \cite{haken} (without the extra variable). That is, one may 
obtain for $G(s,t)$ 
\begin{align}  & G(s,t)=G(t)+G(s)=
\sum_{\lambda}
\frac{g^{em}_{\lambda}\cdot ({g^{em}_{\lambda}})\dagger \cdot
t}{i\bigl(\epsilon_1+w_{\lambda}-
\epsilon_2\bigr)}
+ \label{e66} \\ & +
\sum_{\lambda_s}
\frac{g^{em}_{\lambda_s}\cdot ({g^{em}_{\lambda_s}})\dagger \cdot s}
{i\biggl(\epsilon_1-\epsilon_2+w_{\lambda}+
i\delta\bigl(\epsilon_2+\epsilon_1+w_{\lambda}\bigr)\biggr)} 
\nonumber \end{align}
Substituting from the last equation in Eq (\ref{e60}) 
 one obtains \begin{eqnarray}  && 
P^I(\psi,t,s|\psi^{(0)},0,0)=P^I(\psi^{(0)},0,0)(1+G(t,s))= \label{e67} \\ && =
 P^I(\psi^{(0)},0,0)
\biggl(1+it\Delta
\epsilon_{\lambda}+is\Delta \epsilon_{\lambda_s}\biggr),  \nonumber 
\end{eqnarray} 
where  $\Delta \epsilon_{\lambda}$ and $\Delta \epsilon_{\lambda_s}$ are
\begin{equation} \label{e68} \Delta \epsilon_{\lambda}=\sum_{\lambda_s}
\frac{g^{em}_{\lambda_s}\cdot g^{em}_{\lambda_s}}{\epsilon_2-\epsilon_1-w_{\lambda}}, \ \ \ 
\Delta \epsilon_{\lambda_s}=\sum_{\lambda_s}\frac
{g^{em}_{\lambda_s}\cdot g^{em}_{\lambda_s}}{\epsilon_2-\epsilon_1-w_{\lambda}-
i\delta(\epsilon_2+\epsilon_1+w_{\lambda})}
\end{equation} 
The result in Eq (\ref{e67}) is only for the first-order term in Eq (\ref{e49})
which involves  one emission and one reabsorption done over the intervals
$s_{(0)},s)$, $t_{(0)},t)$. If these emission and reabsorption are repeated for
each one of the many subintervals into which the former finite $s$ and $t$ 
intervals were
subdivided so that 
 all the higher order terms of this process ($N \to \infty$) are taken
into account  one obtains, analogously to the quantum analog \cite{haken} (in which the
variable $s$ is absent), the result \begin{eqnarray} && 
P^I(\psi,t,s|\psi^{(0)},0,0)=
P^I(\psi^{(0)},0,0)(1+G(t,s))=P^I(\psi^{(0)},0,0)\biggl\{1+
\biggl(it\Delta \epsilon_{\lambda}+ \nonumber \\
&&+ \frac{1}{2!}
(it\Delta \epsilon_{\lambda})^2+ \ldots +  \frac{1}{k!}
(it\Delta \epsilon_{\lambda})^k+\ldots\biggr)+ \biggl(is\Delta \epsilon_{\lambda_s}+
\frac{1}{2!}
(it\Delta \epsilon_{\lambda_s})^2+ \ldots  \label{e69} \\ && \ldots + 
\frac{1}{k!}
(it\Delta \epsilon_{\lambda_s})^k+\ldots\biggr)\biggr\} 
 =P^I(\psi^{(0)},0,0)\biggl(e^{it\Delta \epsilon_{\lambda}}
 +e^{is\Delta \epsilon_{\lambda_s}} -1\biggr)
\nonumber \end{eqnarray} 
The left  hand side of Figure 1 shows a Feynman diagram
\cite{haken,mattuck,feynman}  of the emission and
reabsorption process performed once over the relevant $t$ interval whereas 
the right hand side 
of it shows a Feynman diagram
 of the fourth order term 
of this process over the same $t$ interval. 
Now, as required by the SQ theory, the  stationary situations are obtained in 
the limit
of eliminating the extra variable $s$ which is done by equating all the $s$
values to each other and taking to infinity.  Thus,  since, as remarked,  
we have  
 equated the initial $s_{(0)}$ to zero we must  equate all the
other $s$ values to zero. That is, the stationary state is \begin{eqnarray}
&&  \lim_{s\to 0}P^I(\psi,t,s|\psi^{(0)},0,0)= 
\lim_{s\to 0}P^I(\psi^{(0)},0,0)
\biggl(e^{it\Delta \epsilon_{\lambda}}+e^{is\Delta \epsilon_{\lambda_s}}-1\biggr)=
 \label{e70}    \\ && = P^I(\psi^{(0)},0,0)e^{it\Delta \epsilon_{\lambda}} 
 \nonumber \end{eqnarray}
The last result is the one obtained in quantum field theory \cite{haken} 
for  the
same interaction (without any extra variable). The
quantity $\Delta \epsilon_{\lambda}$,  given by the first of Eqs (\ref{e68}), 
has the same form also in the quantum version \cite{haken},   
 where it is 
termed the energy
shift.  This   shift have  
been experimentally demonstrated in the quantum
field theory for the  case of a real
many-state  particle in the famous lamb shift of the Hydrogen atom  
\cite{lamb,haken}. \par 
 Note that,     as
for the gravitational brainwaves  case,  
introducing the expression of the detailed
electron-photon interaction for all the  
  subintervals of $s$ and $t$ of all the stochastic processes 
yields a correlation among them which truly represents, in the stationary
situation,  the corelation of the
real interaction.   That is, when 
   all the values of $s$ are equated to each other  and eliminated 
 the equilibrium stage  is obtained. One  may, also,  note that the 
 elimination of
 the $s$ variable is fulfilled by only equating all its values to each other
 without having to take the infinity limit  (see the discussion before Eq
 (\ref{$A_{14}$}) in Appendix $A$). 
      \par 
      
      \markright{CONCLUDING REMARKS} 

   \protect \section*{\bf Concluding Remarks \label{sec4}} \noindent
   For the first half of this work we have used the fact that the ionic currents
   and charges in 
   cerebral system  radiates electric waves as may be realized by attaching
   electrodes to the scalp. That is, one may physically and 
   logically assume that just as these ionic
   currents and charges in the brain give rise to 
     electric waves so the masses related to these ions and charges 
     should give rise, according to the Einstein's field equations,  
     to  weak
     GW's.   From this 
   we have proceeded to calculate the correlation among an $n$ brain ensemble in
   the sense of finding them at some time  radiating a similar gravitational
   waves if they were found at an earlier time radiating other GW's. We have
   used as a specific example of gravitational wave the cylindrical one which
   have been investigated in a thorough and intensive way (see, for example
   \cite{kuchar}). \par 
   The applied mathematical model, used for calculating the mentioned
   correlation,   was the Parisi-Wu-Namiki SQ theory  \cite{namiki}   which
   assumes a stochastic process performed in an extra dimension  so that 
   at the limit
   of eliminating the relevant extra variable one obtains the physical
   stationary state.  The  hypothetical stochastic process, which is  
   governed by either the Langevin or the
   Fokker-Plank equation,  
    allows  a large ensemble of $n$ 
 different
 variables $\psi$ which  describes this process \cite{parisi,namiki} and 
  represent the mentioned gravitational brainwaves radiated by the $n$
 brain ensemble.   Thus, we have 
   calculated   the correlation in the extra dimension among the 
    $n$ brain ensemble
   and  show that  
   at the limits of (1) eliminating the relevant 
   extra variable and  (2)   maximum correlation one obtains the 
   expected result of finding all of them radiating the same cylindrical GW. \par
   A similar and parallel discussion of the  electron-photon
   interaction, which results in the known Lamb shift, was  carried in the
   second half of this work. This physical example 
    is known to have originated from vacuum fluctuations
   and is in effect one of the first phenomena which were found to be related to
   these fluctuations. Thus, it seems natural to discuss it in terms of the SQ
   theory in which, as mentioned, some stochastic random forces at an extra
   dimension generate at the
   equilibrium stage 
   the known 
   physical stationary state.      
 \par 
 As mentioned, the mechanism which allows the reduction of the random
    stochastic process in the extra dimension to the known physical stationary
 state is the  introduction of this same state in 
 all the $N$ subintervals of all 
the $n$ variables. This  means that once  all 
the different $s$ values are eliminated for  all the subintervals of all the
variables one remains with the same introduced physical stationary state  for
all of them.  The same mechanism  may be shown to   take effect  not only 
for  the
assumed weak cylindrical GW's radiated by the brain and the quantum
fluctuations of the 
Lamb shift discussed here but also for any other physical phenomena which may be
discussed by variational methods.

  \begin{appendix}
  
  \markright{APPENDIX  A:  REPRESENTATION OF THE PARISI-WU-NAMIKI...}
  
\protect \section{APPENDIX A\\ Representation of the Parisi-Wu-Namiki stochastic 
 quantization}

 The Parisi-Wu-Namiki SQ theory \cite{parisi,namiki} for any stochastic process
 \cite{kannan} may use either the
 Langevin equation  \cite{coffey} or the Fokker-Plank one \cite{risken} as its
 basic starting point. For the following introductory representation  of the SQ 
 theory and in Sections II-IV we 
 find it convenient to use the Langevin equation whereas in Sections V-VI 
 we discuss
 the  electron-photon interaction which   results in
 the known Lamb shift \cite{lamb} from the point of view of the Fokker-Plank
 equation. The stochastic process,  which is assumed in the SQ theory  
  to occur in some
 extra dimension $s$, is generally considered to be of the Wienner-Markoff 
 type \cite{kannan} and to be described by the $n$ variables
 $\psi(s,t)=\biggl(\psi_1(s,t),\psi_2(s,t),\ldots \psi_{(n-1)}(s,t),
 \psi_n(s,t)\biggr)$. This stochastic process 
  is also characterized by the $n$
 random forces $\eta(s,t)=\biggl(\eta_1(s,t),\eta_2(s,t),\ldots 
 \eta_{(n-1)}(s,t), \\
 \eta_n(s,t)\biggr)$ which are Gaussian white noise \cite{kannan}. 
   Thus,  denoting the process related to  the    
      $i$ variable   by $\psi_i$, where $ 1 \le i \le n$, 
      one  
   may 
    analyze it   by taking its rate of
    change with respect to $s$ according to  the generalized
Langevin equation \cite{coffey}   
\begin{equation}  \tag{$A_{1}$} \label{$A_{1}$}
\frac{\partial \psi_i(s,t,r)}{\partial s}=K_i(\psi (s,t,r))+\eta_i (s,t,r), 
 \ \ \ \ \
\ \ i=1, 2, \ldots   n, 
\end{equation}
where $n$ denotes the remarked $n$-member ensemble of variables and   
  $\eta_i$  denotes
stochatic process related to  the  variable $\psi_i$.  
      The variables $\psi_i$   depends    
 upon  $s$ and upon the spatial variable $r$   and the time $t$.    
 The
 $K_i$ are given in the SQ theory by 
    \cite{parisi,namiki} \begin{equation} \tag{$A_{2}$} \label{$A_{2}$}
K_i(\psi(s,t,r))=-(\frac{\delta S_i[\psi]}{\delta \psi})_{\psi=\psi(s,t,r)},   \end{equation}
where $S_i$ are  the actions $S_i=\int \int drdtL_i(\psi,\dot \psi)$ 
 and $L_i$ are the Lagrangians. 
For properly discussing the ``evolution'' of the related   process  
 $\psi_i$  one, generally,  subdivides 
 the  $t$ and $s$ intervals $(t_{(0)},t)$,  $(s_{(0)},s)$ 
 into $N$ subintervals $(t_{(0)},t_1)$, $(t_1,t_2)$, \ldots
$(t_{N-1},t)$
  and   $(s_{(0)},s_1)$, $(s_1,s_2)$, \ldots $(s_{N-1},s)$. 
  We assume that the
Langevin Eq (\ref{$A_{1}$}) is satisfied for each member of the ensemble of
variables at each 
subinterval with the following Gaussian constraints \cite{namiki}
 \begin{equation} \tag{$A_{3}$} \label{$A_{3}$} <\!\eta^{(k)}_i(t_k,s_k)\!>=0, \ \ \ 
 <\!\eta^{(k)}_i(t_k,s_k)\eta^{(k)}_j(\grave t_k,\grave s_k)\!>=
2\alpha \delta_{ij}\delta (t_k-\grave t_k)\delta (s_k-\grave s_k), \end{equation} 
where the angular brackets denote an ensemble average with the Gaussian 
distribution, the $k$ superscript denotes the $k$ subinterval from the $N$
available and 
the $i$, $j$ refer to  
  the mentioned $n$ variables where $n \geq i ,  j \geq 1$. 
 Note 
that both intervals $(t_{(0)},t)$,  $(s_{(0)},s)$ of each one of the $n$ variables  
are subdivided, as mentioned,   into  $N$ subintervals.    
The $\alpha$  from Eq (\ref{$A_{3}$}) have different
meanings which depend upon the involved  process and the context in which 
Eqs (\ref{$A_{1}$}) and (\ref{$A_{3}$})
are used. Thus, in the classical regime $\alpha$ is \cite{namiki} 
$\alpha=\frac{k_{\beta}T}{f}$, where $k_{\beta}$, $T$, and $f$ are respectively
the Boltzman constant, the temperature in Kelvin units and the relevant friction
coefficient. In the quantum regime $\alpha$ is identified \cite{namiki} with the
Plank constant $\hbar$.   
  We note that  using Eqs (\ref{$A_{1}$})-(\ref{$A_{3}$}) enables 
  one \cite{namiki} to  discuss 
 a large number of different classical 
and quantum phenomena. 
  It has been shown  
\cite{namiki} that  
  the right hand side 
  of  Eq (\ref{$A_{3}$}) may be derived from the following Gaussian distribution 
  law
  \cite{namiki}     
\begin{equation} \tag{$A_{4}$} \label{$A_{4}$} P_{i}(y)dy_i=
\frac{1}{\sqrt{2\pi(<\!\eta^{(k)}_i\!>)^2}}
\exp(-\frac{(y^{(k)}_i)^2}{2(<\!\eta^{(k)}_i\!>)^2})dy_i, \end{equation} 
which is the probability density for the  variable $\psi_i$ and 
for the subintervals 
 $(s_{(k-1)},s_k)$, $(t_{(k-1)},t_k)$  to have a value 
of $\eta^{(k)}_i$ in $(y^{(k)}_i,y^{(k)}_i+dy_i)$
\cite{namiki}, where 
\begin{equation} \tag{$A_{5}$} \label{$A_{5}$} y^{(k)}_i=\frac{\partial \psi^{(k)}_i(s,t,x)}{\partial
s}-K_i(\psi^{(k)}_i(s,t,x)) \end{equation} 
 
 For the  $n$ variables one may write Eq (\ref{$A_{4}$}) for the subintervals 
 $(s_{(k-1)},s_k)$, $(t_{(k-1)},t_k)$ as 
 \begin{equation} \tag{$A_{6}$} \label{$A_{6}$} P_{ij...}(y)dy=
\exp(-\sum_{i=1}^n\frac{(y^{(k)}_i)^2}{2(<\!\eta^{(k)}_i\!>)^2})
\prod_{i=1}^n\frac{dy_i}{\sqrt{2\pi(<\!\eta^{(k)}_i\!>)^2}}, \end{equation} 
which is the probability density for the $n$ variables $\psi_i \ \ 
 1 \leq i  \leq n$ to have a value 
of $\eta^{(k)}_i$ in $(y^{(k)}_i,y^{(k)}_i+dy_i)$ where $dy=\prod_i dy_i$. The
angular brackets are product over any two variables 
as given in Eq (\ref{$A_{3}$}).  We note in this context that the general
correlation $<\!\eta_i\eta_j\ldots \eta_m\eta_n\!>$ is expressed in terms of
$<\!\eta_i\eta_j\!>$ by \cite{namiki}
\begin{align} & \tag{$A_{7}$} \label{$A_{7}$} <\!\eta_i\eta_j\ldots \eta_m\eta_n\!>=  
 \left\{ \begin{array}{ll} 0 &  {\rm for \ odd \ number \ of \eta's} \\ 
 \sum<\!\eta_i\eta_j\!><\!\eta_m\eta_n\!>\ldots & 
 {\rm for \ even \ number  \ of  \ \eta's }  
  \end{array} \right.   \nonumber   
 \end{align}
 where the sum is taken over every possible pair of $\eta's$.   
For the whole intervals $(s_{(0)},s)$, $(t_{(0)},t)$, which as mentioned were
each subdivided into $N$  subintervals, one may generalize Eq (\ref{$A_{6}$}) as 
\begin{equation} \tag{$A_{8}$} \label{$A_{8}$} P_{ij...}(y)dy=
\exp(-\sum_{i=1}^n\sum_{k=1}^N\frac{(y^{(k)}_i)^2}{2(<\!\eta^{(k)}_i\!>)^2})
\prod_{i=1}^n\prod_{k=1}^N\frac{dy^{(k)}_i}{\sqrt{2\pi(<\!\eta_i^{(k)}\!>)^2}}, 
\end{equation} 
where now the $dy$ at the left is $dy=\prod_i\prod_k dy^{(k)}_i$.  Note that Eqs
(\ref{$A_{4}$}), (\ref{$A_{6}$}) and (\ref{$A_{8}$})  denote probability
densities as realized from the $dy$ at the left hand sides of these equations. 
In order to find the probabilities themselves one have to integrate the right
hand sides of these equations over the appropriate variables. Thus,  using 
Eqs (\ref{$A_{1}$}), (\ref{$A_{3}$}) and (\ref{$A_{5}$}) one may write 
 Eq (\ref{$A_{4}$}) in a
more informative  way as 
\begin{align} &  P_{i}\bigl(\psi^{(k)}_i,t_k,s_k|\psi^{(k-1)}_i,
t_{(k-1)},s_{(k-1)}\bigr)=
\tag{$A_{9}$} \label{$A_{9}$} \\ & =\int d\psi_i^{(k)}\frac{1}{\sqrt{2\pi(2\alpha )}} 
\exp\biggl\{-\frac{\biggl(\frac{\psi^{(k)}_i-\psi^{(k-1)}_i}{(s_k-s_{(k-1)})}
-K_i(\psi_i^{(k-1)})\biggr)^2}{2(2\alpha)}\biggr\}dy, \nonumber
    \end{align} 
 where we have approximated $\frac{\partial \psi^{(k)}_i(s,t,x)}{\partial
s} \approx \frac{\psi_i^{(k)}-\psi_i^{(k-1)}}{(s_k-s_{(k-1)})} $. 
 The $P_i\bigl(\psi_i^{(k)},t_k,s_k|\psi_i^{(k-1)},t_{(k-1)},s_{(k-1)}\bigr)$ of 
 Eq (\ref{$A_{9}$}) 
  is the conditional probability to find the   
  variable $\psi_i$  at
 $t_k$ and $s_k$ with the configuration $\psi^{(k)}_i$ if at $t_{(k-1)}$ 
 and $s_{(k-1)}$ it 
  has the configuration $\psi^{(k-1)}_i$.  Since it involves the same variable
  it may be  termed autocorrelation of $\psi_{(i)}$ over the subintervals
  $(s_{(k-1)},s_{(k)})$, $(t_{(k-1)},t_{(k)})$. In a similar manner one may write
 Eq (\ref{$A_{9}$}) for the whole   ensemble of $n$ variables 
 in the subintervals
 $(s_{(k-1)},s_k)$ and $(t_{(k-1)},t_k)$  as 
 \begin{align} & P_{ij...}\bigl(\psi_{(n)}^{(k)},t_k,s_k|\psi_{(0)}^{(k-1)},
 t_{(k-1)},s_{(k-1)}\bigr)= 
\int \cdots \int \exp\biggl\{-\sum_i\frac{\biggl(\frac{\psi_i^{(k)}-\psi_i^{(k-1)}}{(s_k-s_{(k-1)})}
-K_i(\psi_i^{(k-1)})\biggr)^2}{2(2\alpha)}\biggr\} \cdot \nonumber \\ &
\cdot \prod_{(i=1)}^{(i=n)}\frac{d\psi_{(i)}^{(k)}}{\sqrt{2\pi(2\alpha )}} 
 \tag{$A_{10}$} 
\label{$A_{10}$}    \end{align} 
 And 
  the conditional probability over the whole intervals 
   $(s_{(0)},s)$ and $(t_{(0)},t)$ may  similarly be obtained 
    by adding other
    factors and sums  over the remaining $(N-1)$ subintervals. 
    If one assume $N$ to be  very
    large, and therefore the length of each subinterval to be very small, 
     one may use Feynman path integral \cite{feynman} as follows 
    \begin{align} & P_{ij...}\bigl(\psi,t,s|\psi_{(0)},
 t_{(0)},s_{(0)}\bigr)= \lim_{N \to \infty}C \int \ldots \int
\exp\biggl\{-\sum_{i=1}^n\sum_{k=1}^N
\frac{\biggl(\frac{\psi_i^{(k)}-
\psi_i^{(k-1)}}{(s_k-s_{(k-1)})}
-K_i(\psi_i^{(k-1)})\biggr)^2}{2(2\alpha)}\biggr\} \cdot  \nonumber \\ &
\cdot \prod_{i=1}^n\prod_{k=1}^N\biggl(\frac{d\psi_{(i)}^{(k)}}
{\sqrt{2\pi(2\alpha )}}\biggr),
  \tag{$A_{11}$} 
\label{$A_{11}$}  \end{align} where $C$ is a normalization constant.
   The former formula may equivalently be written as 
   \cite{namiki} \begin{align}  
& P\bigl(\psi,t,s|\psi_{(0)},t_{(0)},s_{0}\bigr)=C\int \cdots \int \cdots 
\int P\bigl(\psi_{(n)}^{N},t_N,s_N|\psi_{(0)}^{(N-1)},t_{(N-1)},s_{(N-1)}\bigr) \cdots 
\tag{$A_{12}$} \label{$A_{12}$} 
 \\ & \cdots 
P\bigl(\psi_{(n)}^k,t_k,s_k|\psi_{(0)}^{k-1},t_{(k-1)},s_{(k-1)}\bigr) 
 \cdots P\bigl(\psi_{(n)}^1,t_1,s_1|\psi_{(0)}^{0},t_{(0)},s_{0}\bigr)d\psi^{N}\cdots d\psi^k 
 \cdots d\psi^1,  \nonumber   \end{align}
where each $P$ at the right is essentially of the form  of Eq (\ref{$A_{10}$})
and the integrals are over the $N$ subintervals.  The last equation,   which is the conditional probability to find the ensemble 
 of $n$ variables  at
 $t$ and $s$ with the configuration $\psi$ if at $t_{(0)}$ and $s_{(0)}$
 they have the configuration $\psi_{(0)}$, is also equivalent \cite{namiki} to    
 the Green's functions $\Delta_{ij\ldots}(t_{(0)},s_{(0)},t_1,s_1, \ldots)$ 
which  determine the correlation among the members of the ensemble 
 \cite{namiki}.   This function, as defined in   \cite{namiki}, is  
 \begin{align} & \Delta_{ij\ldots}(t_{(0)},s_{(0)},t_1,s_1,\ldots)=
 <\!\psi_i(t_{(0)},s_{(0)})\psi_j(t_1,s_1)
 \ldots\!>= \tag{$A_{13}$} \label{$A_{13}$} \\ & = C\int D\psi(t,s)\psi_i(t_{(0)},s_{(0)})\psi_j(t_1,s_1)
 \ldots \exp(-\frac{S_i(\psi(t,s))}{\alpha}),  \nonumber
 \end{align} 
 where $S_i$ are  the actions $S_i=\int ds L_i(\psi,\dot \psi)$, $C$ is a
 normalization constant,  and 
$D\psi(t,s)=\prod_{i=1}^{i=n}d\psi_i(t,s)$. As seen from the last equation the 
$\Delta_{ij\ldots}(t_{(0)},s_{(0)},t_1,s_1 \ldots)$ were expressed as path integrals
\cite{feynman} where the quantum feynman measure $e^{\frac{iS(\psi)}{\hbar}}$ 
is replaced in
 Eq (\ref{$A_{13}$}) and in the following Eq (\ref{$A_{14}$})  
 by $e^{-\frac{S(q)}{\alpha}}$ as required for the classical path integrals
 \cite{namiki,roepstorff}.   \par
  It can be seen that when the $s$'s are different for    the members of 
  the ensemble
so that each have its specific $S_i(\psi(s_i,t))$, $K_i(\psi(s_i,t))$, 
and $\eta_i(s_i,t)$
 the correlation in  (\ref{$A_{13}$}) is obviously zero. 
Thus, in order to have a nonzero value for the probability to find a large part
of the ensemble of variables  having  the same or similar forms  
   we have to consider
the stationary configuration where, as remarked, all the $s$ values are equated
to each other and taken to infinity. For that  matter we take  account of 
the fact 
that the dependence upon $s$ and $t$ is through $\psi$ so  this ensures
\cite{namiki} that this dependence is expressed through the $s$ and $t$
differences. For example, referring to the members $i$ and $j$ the correlation
between them is $\Delta_{ij}(t_i-t_j,s_i-s_j)$, so that for eliminating the $s$
variable from the correlation function one equates all these different $s$'s to
each other.   We, thus,  obtain the following stationary equilibrium correlation 
\cite{namiki} 
\begin{align}  & \Delta_{ij\ldots}(t_{(0)},s_{(0)},\ldots)_{st}=
 <\!\psi_i(t_{(0)},s_{(0)})\psi_j(t_1,s_1)
 \ldots\!>_{st}=C\int D\psi(t)\psi_i(t_{(0)})\psi_j(t_1) \ldots \tag{$A_{14}$}
 \label{$A_{14}$} \\   
 & \ldots \exp(-\frac{S(\psi)}{\alpha}),  \nonumber 
 \end{align}
 where the suffix of $st$  denotes the stationary configuration.  
In other words, the equilibrium   correlation in our case is obtained 
when all the different $s$ values 
   are equated to each other and taken to infinity 
 in which case one remains
with the  known stationary result. 
 \par
 Thus, if all  the    members of the
ensemble of variables   have    
  similar
actions $S$ (in which the $s$ values   
 are equated to each other)  one finds 
with a large probability these  members,   
 in the later equilibrium stage,   with    the
same  result.     That is,  introducing the same similar actions into the
corresponding path integrals one finds this mentioned large probability. 
This has been expicitly shown in Section IV for   
 the cylindrical gravitational  wave
and in Sections  V-VI  
for the Lamb shift case.  

 \end{appendix} 
  
  \begin{appendix}
  
  \markright{APPENDIX  B:  DERIVATION OF THE CORRELATION EXPRESSION....}
  
\protect \section{APPENDIX B\\  Derivation of the correlation expression from 
 Eq (\ref{e39})}

We, now, derive the expression for the correlation  from Eq (\ref{e39}).
For that we may use Eq (\ref{$A_{12}$}) of Appendix $A$ in which we 
substistute for the $P$'s 
from Eqs (\ref{$A_{9}$})-(\ref{$A_{10}$}). As noted in Appendix $A$ the
correlation is calculated not only among  the ensemble of $n$ variables  but also
for each of the $N$ subintervals into which the finite $t$ and $s$ intervals are
divided. Thus, assuming, as noted in Appendix $A$, that $N$ is very large 
we  may use the Feynman path integral of Eq (\ref{$A_{11}$}) and 
write  this correlation  as 
\begin{align} & P_{ij...}\bigl(\psi,t,s|\psi_{(0)},
 t_{(0)},s_{(0)}\bigr)= 
C\int_{-\infty}^{\infty}\ldots \int_{-\infty}^{\infty}\ldots
\int_{-\infty}^{\infty}\exp\biggl\{-\sum_{k=1}^N\sum_{i=1}^n
\frac{1}{4\alpha(s_{k}-s_{(k-1)})^2}\biggl(\psi_i^{(k)}- \nonumber \\ & -
\psi_{(i-1)}^{(k)}  
-K_i(\psi_{(i-1)}^{(k)})(s_{k}-s_{(k-1)})\biggr)^2\biggr\} 
\prod_{k=1}^{k=N}\prod_{i=1}^{i=(n-1)} \frac{d\psi^k_i}
{\sqrt{2\pi\bigl(2\alpha \bigr)}}   
 \tag{$B_{1}$} 
\label{$B_{1}$}   \end{align}  
where $C$ is a normalization constant to be determined later from $\int 
P_{ij...}\bigl(\psi,t,s|\psi^0,
 t_{(0)},s_{(0)}\bigr)d\psi=1$. Note that in the exponent of 
 Eq (\ref{$B_{1}$}), in contrast to that of Eq
 (\ref{$A_{11}$}) in Appendix $A$, the sum over $i$ precedes that over $k$ and,
 therefore, the squared expression involves the variables $\psi_{(i)}^{(k)}$, 
 $\psi_{(i-1)}^{(k)}$  etc   (instead of $\psi_{(i)}^{(k)}$, 
 $\psi_{(i)}^{(k-1)}$  of (\ref{$A_{11}$})).     Note also that the number of integrals are 
 $N \times (n-1)$ over the $N$ 
subintervals and $(n-1)$
variables  which is related to the fact that the suffix $i$  in the exponent 
 is summed from $i=1$ to $i=n$  whereas the $i$ in the differentials outside the
 exponent is summed up to $i=n-1$ 
 (compare with equation (4.4 in \cite{namiki}). The reason for this 
is that each $\psi_{(i)}^{(k)}$, except for $i=0$ and $i=n$,  with 
superscript  $k$ and suffix $i$ appears in two 
consecutive
squared 
expressions of 
  the sum over $i$ so for calculating  the correlation 
for the observer $i$ over the 
 subinterval
$(s_k-s_{(k-1)})$ one has   to  solve  
  the following integral which is 
related to $\psi_{(i)}^{(k)}$.   \begin{align} &
P_{i}\bigl(\psi^{(k)}_i,t_{(k)},s_{(k)}|\psi^{(k)}_{(i-1)},t_{(k-1)},s_{(k-1)}\bigr)=
 \nonumber \\ & =
\int_{-\infty}^{\infty}\exp\biggl\{-
\biggl[\frac{\biggl(\psi_i^{(k)}-
\psi_{(i-1)}^{(k)}
-K_i(\psi_{(i-1)}^{(k)})(s_k-s_{(k-1)})\biggr)^2}{2(2\alpha)
\bigl(s_k-s_{(k-1)}\bigr)^2}
+ \tag{$B_{2}$} 
\label{$B_{2}$}  \\ & + \frac{\biggl(\psi_{(i+1)}^{(k)}-
\psi_i^{(k)}
-K_i(\psi_i^{(k)})(s_{(k)}-s_{(k-1)})\biggr)^2}{2(2\alpha)
\bigl(s_{(k)}-s_{(k-1)}\bigr)^2} \biggr]\biggr\} 
 \frac{d\psi^{(k)}_i}{\sqrt{2\pi(2\alpha)}}  
\nonumber  \end{align}
The solution of this integral involves the substitution  
 for $K_i(\psi_{(i-1)}^{(k)})$ and 
$K_i(\psi_i^{(k)})$ from Eqs (\ref{e32}) and   (\ref{e36}) so that one may write 
the two  squared expressions   of Eq (\ref{$B_{2}$}) as
\begin{align} & \frac{\biggl(\psi_i^{(k)}-
\psi_{(i-1)}^{(k)}
-K_i(\psi_{(i-1)}^{(k)})(s_k-s_{(k-1)})^2\biggr)^2}{2(2\alpha)
\bigl(s_{(k)}-s_{(k-1)}\bigr)}= \frac{1}{2(2\alpha)
\bigl(s_{(k)}-s_{(k-1)}\bigr)^2}\biggl[\psi_i^{(k)}- \nonumber \\ & -
\psi_{(i-1)}^{(k)}-2\pi \biggl(B_1(R,t)\psi^{(k)}_{(i-1)}-
B_2(R,t)-
iB_3(R,t)\biggr)(s_k-s_{(k-1)})\biggr]^2  \tag{$B_{3}$} 
\label{$B_{3}$} \\ & 
\frac{\biggl(\psi_{(i+1)}^{(k)}-
\psi_i^{(k)}
-K_i(\psi_i^{(k)})(s_{(k)}-s_{(k-1)})\biggr)^2}{2(2\alpha)
\bigl(s_{(k)}-s_{(k-1)}\bigr)^2}= \frac{1}{2(2\alpha)
\bigl(s_{(k)}-s_{(k-1)}\bigr)^2}\biggl[\psi_{(i+1)}^{(k)}- \nonumber  \\ & -
\psi_i^{(k)}-2\pi\biggl(B_1(R,t)\psi^{(k)}_i-  
B_2(R,t)-
iB_3(R,t)\biggr)(s_{(k)}-s_{(k-1)})\biggr]^2
\nonumber \end{align}
 In order to deal with manageable expressions we first
assume that in the limit of large $N$ and $n$ the subintervals over $t$ and $s$
are equal so that one may write  for any integral $k$
\begin{align} & \Delta s_k=(s_k-s_{(k-1)})=\Delta s_{(k+1)}=(s_{(k+1)}-s_{(k)})
=\Delta s \tag{$B_{4}$} 
\label{$B_{4}$}  \\ & \Delta t_k=(t_k-t_{(k-1)})=\Delta t_{(k+1)}=(t_{(k+1)}-
t_{(k)})=
\Delta t 
\nonumber 
\end{align}
We, now, define the following expressions 
\begin{equation} \tag{$B_{5}$} 
\label{$B_{5}$} a_1=(1+2\pi B_1\Delta s)^2, \ \ \ \  a_2=2\pi \Delta s(B_2+iB_3) 
\end{equation}
 Using  Eqs (\ref{$B_{3}$})-(\ref{$B_{5}$}) one may write the two squared 
 terms of Eq (\ref{$B_{2}$})   as
 \begin{align} & 
\frac{\biggl(\psi_i^{(k)}-
\psi_{(i-1)}^{(k)}
-K_i(\psi_{(i-1)}^{(k)})(s_k-s_{(k-1)})\biggr)^2}{2(2\alpha)
\bigl(s_{(k)}-s_{(k-1)}\bigr)^2} + \frac{\biggl(\psi_{(i+1)}^{(k)}-
\psi_i^{(k)}
-K_i(\psi_i^{(k)})(s_{(k)}-s_{(k-1)})\biggr)^2}{2(2\alpha)
\bigl(s_{(k)}-s_{(k-1)}\bigr)^2}= \nonumber \\ & =
\frac{1}{4\alpha(\Delta
s)^2}\biggl\{\biggl(\psi^{(k)}_i-\sqrt{a_1}\psi^{(k)}_{(i-1)}+a_2\biggr)^2+
\biggl(\psi^{(k)}_{(i+1)}-\sqrt{a_1}\psi^{(k)}_i+a_2\biggr)^2\biggr\}= 
\tag{$B_{6}$} 
\label{$B_{6}$} \\
& =\frac{1}{4\alpha(\Delta
s)^2}\biggl\{(\psi^k_i)^2+a_1(\psi^{(k)}_{(i-1)})^2-2\sqrt{a_1}\psi^{(k)}_i
\psi^{(k)}_{(i-1)}
+2a_2\psi^{(k)}_i-2a_2\sqrt{a_1}\psi^{(k)}_{(i-1)}+ \nonumber \\ & + 
(\psi^{(k)}_{(i+1)})^2+ 
a_1(\psi^{(k)}_i)^2- 
 2\sqrt{a_1}\psi^{(k)}_{(i+1)}\psi^{(k)}_i+
 2a_2\psi^{(k)}_{(i+1)}-2a_2\sqrt{a_1}\psi^{(k)}_i+2a_2^2\biggr\}
\nonumber \end{align}
The last result is now substituted for the  two squared terms of Eq
(\ref{$B_{2}$}) and the integral over $\psi^k_i$ may be solved by using the
following integral \cite{abramowitz}
\begin{equation} \tag{$B_{7}$} 
\label{$B_{7}$} \int_{-\infty}^{\infty}dx
\exp\bigl(-\bigl(ax^2+bx+c\bigr)\bigr)=\sqrt{\frac{\pi}{a}}
\exp\bigl(\frac{(b^2-4ac)}{4a}\bigr) \end{equation}
Thus,  using Eq (\ref{$B_{6}$}), one may find the   appropriate coefficients 
$a_{\psi^k_i}$,
$b_{\psi^k_i}$ and $c_{\psi^k_i}$, related to $\psi^k_i$,  to 
be substituted in the integral (\ref{$B_{2}$}) as follows
\begin{align} & 
a_{\psi^k_i}=\frac{(1+a_1)}{4\alpha(\Delta s)^2}, \ \ \ \ \
b_{\psi^k_i}=\frac{\biggl(2a_2\bigl(1-\sqrt{a_1}\bigr)-2\sqrt{a_1}
\bigl(\psi^{(k)}_{(i-1)}+
\psi^{(k)}_{(i+1)}\bigr)\biggr)}
{4\alpha(\Delta s)^2} 
\tag{$B_{8}$} 
\label{$B_{8}$} \\ & 
c_{\psi^k_i}=\frac{\biggl[(\psi^{(k)}_{(i+1)})^2+a_1(\psi^{(k)}_{(i-1)})^2+
2a_2\biggl(a_2+\psi^{(k)}_{(i+1)}-\sqrt{a_1}\psi^{(k)}_{(i-1)}\biggr)\biggr]}
{4\alpha(\Delta s)^2}   \nonumber   \end{align}
Using the last expressions for the coefficients $a_{\psi^k_i}$,
$b_{\psi^k_i}$ and $c_{\psi^k_i}$  one may realize, after some calculations, 
  that they satisfy the
following relation
\begin{equation}
\frac{b^2_{\psi^k_i}-4a_{\psi^k_i}c_{\psi^k_i}}{4a_{\psi^k_i}}=
-\frac{1}{4(1+a_1)\alpha(\Delta s)^2}\biggl(\bigl(\psi^{(k)}_{(i+1)}-
a_1\psi^{(k)}_{(i-1)}\bigr)+
a_2\bigl(1+\sqrt{a_1}\bigr)\biggr)^2
\tag{$B_{9}$} 
\label{$B_{9}$} \end{equation}
Thus, using the former discussion and, especially, the integral (\ref{$B_{7}$})
one is able to solve the integral from Eq (\ref{$B_{2}$}) and write it as
\begin{align} &
P_{i}\bigl(\psi^{(k)}_{(i+1)},t_{(k)},s_{(k)}|\psi^{(k)}_{(i-1)},t_{(k-1)},s_{(k-1)}\bigr)=
 \nonumber \\ &
 =\int_{-\infty}^{\infty}\frac{d\psi^{(k)}_i}{\sqrt{2\pi(2\alpha)}}
 \exp\biggl[-
 \biggl(a_{\psi^{(k)}_i}(\psi^{(k)}_i)^2+b_{\psi^{(k)}_i}\psi^{(k)}_i
 +c_{\psi^{(k)}_i}\biggr)\biggl]=
 \frac{1}{\sqrt{4\alpha  a_{\psi^k_i}}} \cdot \tag{$B_{10}$} 
\label{$B_{10}$} \\ & \cdot \exp
 \biggl(\frac{b^2_{\psi^k_i}-4a_{\psi^k_i}c_{\psi^k_i}}{4a_{\psi^k_i}}\biggr)
 = \frac{\Delta s}{\sqrt{(1+a_1)}}\exp\biggl\{-\biggl[\frac{1}{4(1+a_1)
 \alpha(\Delta s)^2}\biggl(\bigl(\psi^{(k)}_{(i+1)}- \nonumber \\ & - 
 a_1\psi^{(k)}_{(i-1)}\bigr)+
a_2\bigl(1+\sqrt{a_1}\bigr)\biggr)^2\biggr]\biggr\}
 \nonumber \end{align}
 The last result is the correlation for the variable $\psi_{(i)}$ over 
 the subinterval
 $(s_k-s_{(k-1)})$ and it means the conditional probability to find this 
 variable
  at $s=s_{(k)}$ and $t=t_{(k)}$ at the state $\psi_{(i+1)}^{(k)}$ 
 if at 
  $s=s_{(k-1)}$ and $t=t_{(k-1)}$ it was at the state $\psi_{(i-1)}^{(k)}$. Note
  that the superscript  of the variable $\psi_{(i-1)}$ at the beginning of the
  subintervals $s_{(k-1)}$ and 
  $t_{(k-1)}$ is the same as that at the end of it, i.e., $k$.  
  If one wish to find the correlation of the two  variables 
  $\psi_{(i)}$ and $\psi_{(i+1)}$ for
 the same subinterval  $\Delta s$  then he has
 to add to the last result another squared term from the general relation 
 (\ref{$B_{1}$})  and
 perform the required integration over $\psi^{(k)}_{(i+1)}$ as follows
\begin{align} &
P_{i,(i+1)}\bigl(\psi^{(k)}_{(i+2)},t_{(k)},s_{(k)}|\psi^{(k)}_{(i-1)},t_{(k-1)},s_{(k-1)}\bigr)=
 \nonumber \\ & =
\frac{\Delta s}{\sqrt{(1+a_1)}}\int_{-\infty}^{\infty}\exp\biggl\{-\biggl[
\frac{1}{4(1+a_1)\alpha(\Delta s)^2}\biggl\{\biggl(\bigl(\psi^{(k)}_{(i+1)}-
a_1\psi^{(k)}_{(i-1)}\bigr)+
a_2\bigl(1+\sqrt{a_1}\bigr)\biggr)^2+ \nonumber \\ & +
 \biggl(\psi_{(i+2)}^{(k)}-
\psi_{(i+1)}^{(k)}
-K_i(\psi_{(i+1)}^{(k)})\Delta s\biggr)^2(1+a_1)\biggr\}
\biggr]\biggr\}   
   \frac{d\psi^{(k)}_{(i+1)}}{\sqrt{2\pi(2\alpha)}}
\tag{$B_{11}$} 
\label{$B_{11}$} \end{align}
In this case the corresponding $a_{\psi^{(k)}_{(i+1)}}$, 
$b_{\psi^{(k)}_{(i+1)}}$ and 
$c_{\psi^{(k)}_{(i+1)}}$ are 
\begin{align} & 
a_{\psi^{(k)}_{(i+1)}}=\frac{\bigl(1+a_1+a^2_1)\bigr)}{2(2\alpha)(\Delta s)^2},
\nonumber \\ & 
b_{\psi^{(k)}_{(i+1)}}=\frac{\biggl(2a_2\bigl(1+\sqrt{a_1}\bigr)-
2a_1\psi^{(k)}_{(i-1)}-(1+a_1)\biggl(2\sqrt{a_1}
\psi^{(k)}_{(i+2)}+2a_2\sqrt{a_1}\biggr)}
{2(2\alpha)(\Delta s)^2} 
\tag{$B_{12}$} 
\label{$B_{12}$} \\ & 
c_{\psi^{(k)}_{(i+1)}}=\frac{\biggl(a_1^2(\psi^{(k)}_{(i-1)})^2+
a_2^2(1+\sqrt{a_1})^2-
2a_1a_2(1+\sqrt{a_1})\psi^{(k)}_{(i-1)}+(1+a_1)\bigl(\psi^{(k)}_{(i+2)}
+a_2\bigr)^2
\biggr)}
{2(2\alpha)(\Delta s)^2}   \nonumber   \end{align}
Thus, using the last equations and the integral from Eq (\ref{$B_{7}$}) one may
write the correlation from Eq (\ref{$B_{11}$}) as
\begin{align} &
P_{i}\bigl(\psi^{(k)}_{(i+2)},t_{(k)},s_{(k)}|\psi^{(k)}_{(i-1)},
t_{(k-1)},s_{(k-1)}\bigr)=
 \tag{$B_{13}$} 
\label{$B_{13}$}  \\ &
 =\frac{(\Delta s)^2}{\sqrt{(1+a_1)\bigl(1+a_1+a_1^2\bigr)}}
 \exp\biggl\{-\biggl[
\frac{\biggl(\psi^{(k)}_{(i+2)}-a_1\sqrt{a_1}\psi^{(k)}_{(i-1)}+
a_2\bigl(1+\sqrt{a_1}+(\sqrt{a_1})^2\bigr)\biggr)^2}
{4\alpha(\Delta s)^2\bigl(1+a_1+a_1^2\bigr)}\biggr]\biggr\}  \nonumber 
  \end{align}
  Using the results of Eq (\ref{$B_{10}$}) for the observer $i$ 
   one may realize that the correlation from Eq
  (\ref{$B_{13}$}) means the conditional probability to find at $s=s_{(k)}$ 
  and $t=t_{(k)}$ the two variables $\psi_{(i)}$ and $\psi_{(i+1)}$ at 
  the respective states of
  $\psi_{(i+1)}^{(k)}$ and   $\psi_{(i+2)}^{(k)}$ if  at $s=s_{(k-1)}$ 
  and $t=t_{(k-1)}$ they  were  at the   states $\psi_{(i-1)}^{(k)}$, 
  $\psi_{(i)}^{(k)}$.  As remarked after Eq (\ref{$B_{10}$}) 
  the superscripts  of the variables  $\psi_{(i-1)}^{(k)}$, 
  $\psi_{(i)}^{(k)}$  at the beginning of the
  subintervals $s_{(k-1)}$ and 
  $t_{(k-1)}$ are the same as that at the end of it, i.e., $k$. 
One may, now, realize that the correlation of the $n$ observers $i,j,l...$ 
over 
 the subinterval 
$(s_{(k-1)},
s_{(k)})$ may be obtained from   the results of 
Eqs (\ref{$B_{10}$}),  (\ref{$B_{13}$}) and  from Eq (\ref{$B_{1}$}) as 
\begin{align} & P_{i,j,l...}\bigl(\psi^{(k)}_{(n)},t_{(k)},s_{(k)}|\psi^k_0,
 t_{(k-1)},s_{(k-1)}\bigr)=   
  \frac{(\Delta s)^{(n-1)}}{\sqrt{\prod_{j=1}^{j=(n-1)}(\sum_{m=0}^{m=j}a_1^m)}} \cdot \nonumber
  \\ & \cdot 
\exp\biggl\{-
\frac{1}{4\alpha (\Delta s)^2\sum_{p=0}^{p=(n-1)}a_1^p}\biggl(\psi_n^{(k)}- 
(\sqrt{a_1})^{n+1} 
\psi_0^{(k)} 
+a_2\sum_{r=0}^{r=n+1}(\sqrt{a_1})^r\biggr)^2\biggr\},   
\tag{$B_{14}$} 
\label{$B_{14}$}  \end{align}  
The last correlation means the conditional probability to  find 
at $s=s_{(k)}$ 
  and $t=t_{(k)}$ the  variables $\psi_{(n-1)}$, $\psi_{(n-2)}, \ \ldots  
  \psi_{(1)}$ at 
  the respective states of
  $\psi_{(n)}^{(k)}$,    $\psi_{(n-1)}^{(k)}, \ \ldots \psi_{(2)}$ if  at 
  $s=s_{(k-1)}$ 
  and $t=t_{(k-1)}$ they were at  
  $\psi_{(n-2)}^{(k)}$,    $\psi_{(n-3)}^{(k)}, \ \ldots \psi^{(k)}_{(0)}$. 
  Note again,  as remarked after Eqs (\ref{$B_{10}$}) and (\ref{$B_{13}$}), that
  the superscripts of each of the $(n-1)$ variables at  the beginning of the
  subintervals $s_{(k-1)}$ and 
  $t_{(k-1)}$ are the same as that at the end of it, i.e., $k$.  
In a similar manner one may calculate, through the double sum  
$\sum_{k=1}^N\sum_{i=1}^n
\frac{1}{4\alpha(s_{k}-s_{(k-1)})^2}\biggl(\psi_i^{(k)}- 
\psi_{(i-1)}^{(k)}  
-K_i(\psi_{(i-1)}^{(k)})(s_{k}-s_{(k-1)})\biggr)^2 $ in the exponent of Eq
(\ref{$B_{1}$}),    
 the correlation for each of the other 
$(N-1)$  subintervals.  Taking into account that all these  subintervals are, 
as realized from Eq (\ref{$B_{4}$}), identical it is obvious 
that  the result of calculating the correlation  for each of
them  is, except for change of the superscripts 
$k$ of
$\psi$,  the same as that of Eq (\ref{$B_{14}$}).  Thus, the correlation of the
ensemble of the $n$  observers over  all the $N$ subintervals 
$(s_{(0)},
s_1),  \ldots (s_{(N-1)},s_{(N)})$ is obtained by multiplying together 
$N$ expressions of the kind of Eq (\ref{$B_{14}$}).  
 That is,   
\begin{align} & P_{i,j,l,....}\bigl(\psi^{(N)}_{(n)},t_{(N)},s_{(N)}|\psi^{(1)}_0,
 t_{(0)},s_{(0)}\bigr)=   
  \frac{C(\Delta s)^{N(n-1)}}{\biggl(\prod_{j=1}^{j=(n-1)}(\sum_{m=0}^{m=j}a_1^m)
  \biggr)^{\frac{N}{2}}}
\cdot \tag{$B_{15}$} 
\label{$B_{15}$}  \\ & \cdot
\exp\biggl\{-
\frac{N}{4\alpha (\Delta s)^2\sum_{k=0}^{k=(n-1)}a_1^k}\biggl(\psi_n^{(N)}- 
 (\sqrt{a_1})^{n+1} 
\psi_0^{(N)} 
+a_2\sum_{r=0}^{r=n+1}(\sqrt{a_1})^r\biggr)^2\biggr\},       
\nonumber  \end{align}   
where $C$ is the normalizing constant which  is, as  mentioned after 
Eq (\ref{$B_{1}$}),  
calculated  from    the normalizing condition
\cite{namiki} 
$\int P_{ij....}\bigl(\psi_{(n-1)},t_{(N)},s_{(N)}|\psi_0,
 t_{(0)},s_{(0)}\bigr)d\psi=1$. Using the results of Eqs 
 (\ref{$B_{10}$}),  (\ref{$B_{13}$})-(\ref{$B_{14}$})  one may realize that 
 the correlation from Eq (\ref{$B_{15}$}) means the
 conditional probability to find  at $s=s_{(N)}$ 
  and $t=t_{(N)}$ the  variables $\psi_{(n-1)}$, $\psi_{(n-2)}, \ \ldots  
  \psi_{(1)}$ at 
  the respective states of
  $\psi_{(n)}^{(N)}$,    $\psi_{(n-1)}^{(N)}, \ \ldots \psi_{(2)}^{(N)}$   if  
   at 
  $s=s_{(N-1)}$ 
  and $t=t_{(N-1)}$ they   were found   at   
  $\psi_{(n-2)}^{(N)}$,    $\psi_{(n-3)}^{(N)}, \ \ldots \psi_{(0)}^{(N)}$
  and at  $s=s_{(N-3)}$ 
  and $t=t_{(N-3)}$ they were found at    
  $\psi_{(n-2)}^{(N-2)}$,    $\psi_{(n-3)}^{(N-2)}, \ \ldots \psi_{(0)}^{(N-2)}$
 $\ldots \ldots$ and at $s=s_{(0)}$ and $t=t_{(0)}$ they were at 
  $\psi_{(n-2)}^{(1)}$,    $\psi_{(n-3)}^{(1)}, \ \ldots \psi_{(0)}^{(1)}$.  
  That is, the conditional probability here involves $N$ conditions at the
  beginnings of the $N$
  subintervals  
 so that, as remarked for the specific cases of Eqs (\ref{$B_{10}$}), 
 (\ref{$B_{13}$}) and (\ref{$B_{14}$}),  
   the superscript  of each  of  the $(n-1)$ ensemble of 
 variables $\psi_{(n-1)}$, $\psi_{(n-2)}, \ \ldots  
  \psi_{(1)}$ at the beginning of each of the $N$ subintervals $(s_{(N-1)},s_{(N)}), 
  (s_{(N-3)},s_{(N-2)}), \ \ldots (s_{(0)},s_{(1)})$ is as same as that at
  end of it.  
 Thus, substituting from Eq (\ref{$B_{15}$}) into this
 normalizing equation one obtains  
 \begin{align} & \int_{-\infty}^{\infty}P_{ij....}\bigl(\psi_{(n)},t_{(N)},s_{(N)}|\psi_0,
 t_{(0)},s_{(0)}\bigr)d\psi^{(N)}_{(n)}=   
  \frac{C(\Delta s)^{N(n-1)}}{\biggl(\prod_{j=1}^{j=(n-1)}(\sum_{m=0}^{m=j}a_1^m)
  \biggr)^{\frac{N}{2}}}
\cdot \tag{$B_{16}$} 
\label{$B_{16}$}  \\ & \dot \int_{-\infty}^{\infty}
\exp\biggl\{-
\frac{N}{4\alpha (\Delta s)^2\sum_{k=0}^{k=(n-1)}a_1^k}\biggl(\psi_n^{(N)}- 
 (\sqrt{a_1})^{n+1} 
\psi_0^{(N)} 
+a_2\sum_{r=0}^{r=n+1}(\sqrt{a_1})^r\biggr)^2\biggr\} \cdot \nonumber \\ &
\cdot  d\psi^N_n =1       
\nonumber  \end{align}   
 Note that the value of $\psi_{(0)}^{(N)}$ 
is generally given  so the variable is $\psi_{(n)}^{(N)}$ as denoted in the last
expression. Now, expanding the
squared expression in the last equation and 
using the integral from Eq (\ref{$B_{7}$}) 
 one may note that the coefficients $a_{\psi_n^k}$, 
$b_{\psi_n^k}$, $c_{\psi_n^k}$ are 
\begin{align} & 
a_{\psi_{(n)}^{(N)}}= \frac{N}{4\alpha (\Delta s)^2\sum_{k=0}^{k=(n-1)}a_1^k} \nonumber
\\ & 
 b_{\psi_{(n)}^{(N)}}=\frac{N\biggl(2a_2\sum_{r=0}^{r=n+1}(\sqrt{a_1})^r-2(\sqrt{a_1})^{n+1} 
\psi_0^{(k)}\biggr)}{4\alpha (\Delta s)^2\sum_{k=0}^{k=(n-1)}a_1^k}  \tag{$B_{17}$} 
\label{$B_{17}$} \\ & 
 c_{\psi_{(n)}^{(N)}}= \frac{N\biggl(\bigl((\sqrt{a_1})^{n+1} \psi_0^{(k)}\bigr)^2+
 \bigl(a_2\sum_{r=0}^{r=n+1}(\sqrt{a_1})^r\bigr)^2-2a_2\sum_{r=0}^{r=n+1}
 (\sqrt{a_1})^r
( \sqrt{a_1})^{(n+1)} \psi_0^{(N)}\biggr)}
{4\alpha (\Delta s)^2\sum_{k=0}^{k=(n-1)}a_1^k}
 \nonumber  \end{align} 
Thus, substituting from the last equations into  Eq (\ref{$B_{7}$}) and noting
that $(b_{\psi_{(n)}^{(N)}})^2- 4a_{\psi_{(n)}^{(N)}} c_{\psi_{(n)}^{(N)}}=0$
one may calculate  the integral from Eq (\ref{$B_{16}$}) over $\psi_{(n)}^{(N)}$
as \begin{align} & 
 \int_{-\infty}^{\infty}
 d\psi_{(n)}^{(N)}
\exp\biggl\{-
\frac{N}{4\alpha (\Delta s)^2\sum_{k=0}^{k=(n-1)}a_1^k} 
\biggl(\psi_{(n)}^{(N)}- (\sqrt{a_1})^{n+1} 
\psi_0^{(N)} + \nonumber \\ & +
a_2\sum_{r=0}^{r=n+1}(\sqrt{a_1})^r\biggr)^2\biggr\} 
= 
\biggl(\frac{4\pi \alpha (\Delta s)^2\sum_{k=0}^{k=(n-1)}a_1^k}{N}\biggr)^{\frac{1}{2}} 
 \tag{$B_{18}$} 
\label{$B_{18}$} 
\end{align}
Substituting the last result into Eq (\ref{$B_{16}$}) and solving for $C$
  one obtains
 \begin{equation} \tag{$B_{19}$} 
\label{$B_{19}$} C=\frac{N^{\frac{1}{2}}\biggl(\prod_{j=1}^{j=(n-1)}(\sum_{m=0}^{m=j}a_1^m)
  \biggr)^{\frac{N}{2}}}
 {(\Delta s)^{N(n-1)}\biggl(4\pi \alpha (\Delta s)^2\sum_{k=0}^{k=(n-1)}a_1^k\biggr)^{\frac{1}{2}}}
 \end{equation} 
 Substituting this value of $C$ in Eq (\ref{$B_{15}$}) one obtains the complete
 expression for the correlation of the $n$ observers over the $N$
 subintervals as written in Eq (\ref{e39}) 
 \begin{align} & P_{i,j,l,.....}\bigl(\psi_{(n)},t_N,s_N|\psi_0,
 t_{(0)},s_{(0)}\bigr)= \biggl(\frac{N}
 {4\pi \alpha (\Delta s)^2\sum_{k=0}^{k=(n-1)}a_1^k}\biggr)^{\frac{1}{2}}
\cdot \tag{$B_{20}$} 
\label{$B_{20}$}  \\ & \cdot
\exp\biggl\{-
\frac{N}{4\alpha (\Delta s)^2\sum_{k=0}^{k=(n-1)}a_1^k}\biggl(\psi_n^{(N)}- 
 (\sqrt{a_1})^{n+1} 
\psi_0^{(N)} 
+a_2\sum_{r=0}^{r=n+1}(\sqrt{a_1})^r\biggr)^2\biggr\}       
\nonumber  \end{align}

 \end{appendix} 
 
 \markright{REFERENCES}
 
\newpage
\bigskip 
\bibliographystyle{}

\begin{figure}
\includegraphics[width=6 in]{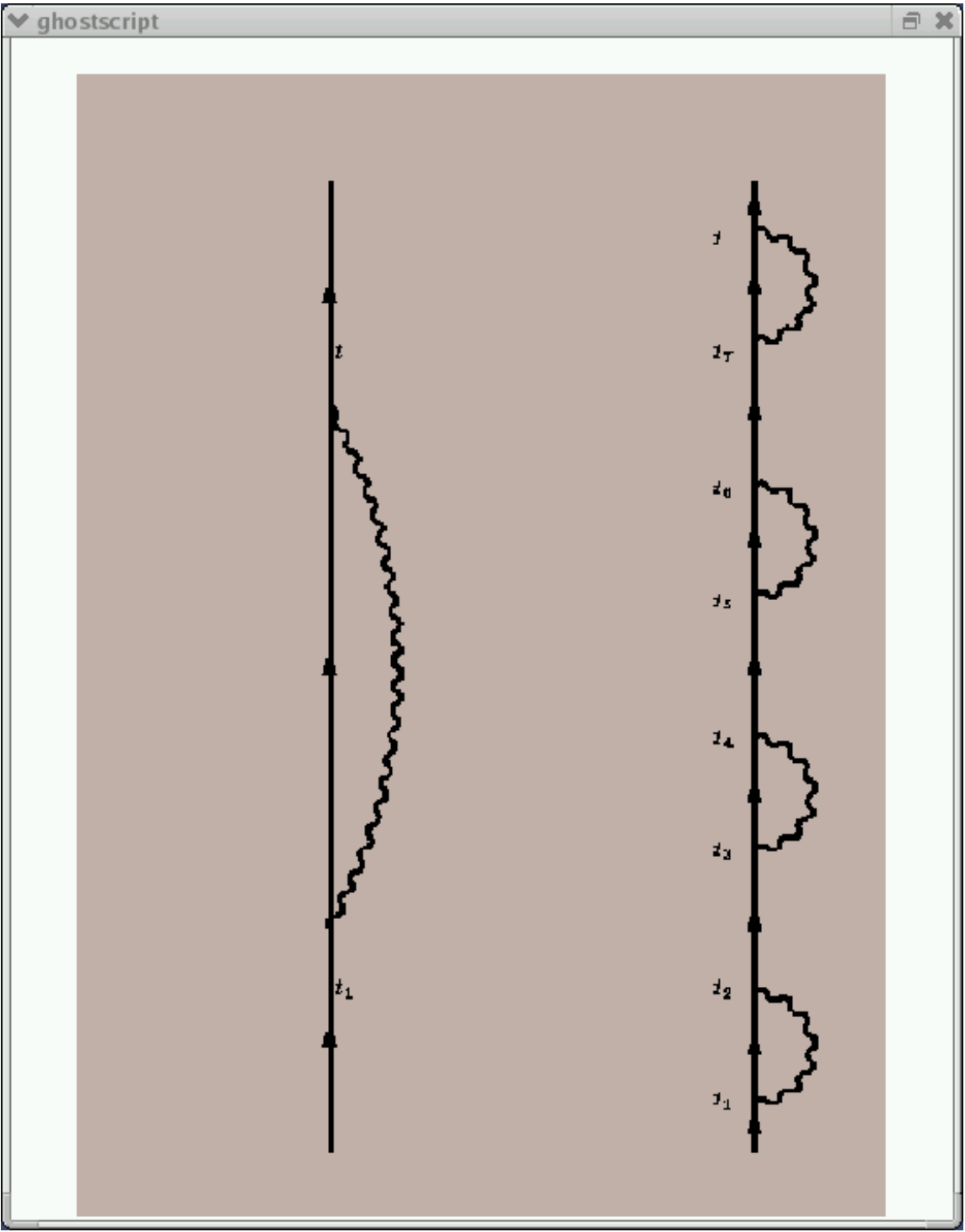}
\caption[fegf3]{The left hand side of the figure shows a Feynman diagram of the 
 process of emitting and reabsorbing a photon in the time interval
$(t_{(0)},t)$ 
where the energy is not conserved. The electron is represented in the figure by
the directed arrow and the photon by the wavy line.  The right hand side of the
figure shows the same  process repeated four times, in a perturbative
manner,  over the same time interval.  }
\end{figure}

   \end{document}